\documentclass[journal]{IEEEtran}

\usepackage{amsmath,amssymb,amsthm}
\usepackage{graphicx}
\usepackage{booktabs}
\usepackage{siunitx}
\usepackage{url}
\usepackage[colorlinks=true,linkcolor=black,citecolor=black,urlcolor=blue]{hyperref}

\DeclareSIUnit{\ohm}{\ensuremath{\Omega}}
\DeclareSIUnit{\dBi}{dBi}

\newtheorem{theorem}{Theorem}
\newtheorem{proposition}{Proposition}
\newtheorem{corollary}{Corollary}
\theoremstyle{definition}
\newtheorem{remark}{Remark}

\DeclareMathOperator{\Rez}{Re}
\DeclareMathOperator{\Imz}{Im}
\newcommand{\Zin}{Z_{\mathrm{in}}}
\newcommand{\Zres}{Z_{\mathrm{res}}}
\newcommand{\Zact}{Z_{\mathrm{act}}}

\newcommand{\etamm}{\eta_{\mathrm{mm}}}
\newcommand{\etarad}{\eta_{\mathrm{rad}}}
\newcommand{\ye}{y_{e}}

\begin{document}

\title{Feasibility and Convex Design of Probe-Position Matching\\in a
Scanning X-Band Radar Array}

\author{William~Khalili~Jr.%
\thanks{The author is an independent researcher in radar signal processing
(e-mail: william.khalilijr@gmail.com).}%
\thanks{\emph{Data availability:} the solver exports, the analysis scripts
that generate every figure and table, and the solver macros that build the
models of Table~\ref{tab:solver} are provided as supplementary material.}}

\markboth{IEEE Transactions on Antennas and Propagation}%
{Khalili: Probe-Position Matching in a Scanning X-Band Radar Array}

\maketitle

\begin{abstract}
The probe position of a microstrip element is normally chosen from the
isolated-element input resistance and verified, if at all, by simulating the
array. This paper replaces that procedure with a decision available before
any array optimisation is attempted. A single Floquet unit-cell solution at
one arbitrary probe position is decomposed into a feed inductance, a
transformer ratio carrying the probe position, and an array-loaded resonator
which, within that decomposition and under its stated hypotheses, is common
to every candidate probe position. Three closed-form results follow. The set
of input impedances reachable by probe position and resonant length is a disk
in the impedance plane; an exact match to a real reference $Z_{0}$ therefore
exists if and only if $R_{p}\geq Z_{0}+X_{p}^{2}/Z_{0}$, a criterion strictly
stronger than the familiar requirement that the transformed resistance exceed
twice the feed reactance; and when it fails, the best attainable reflection
follows from the image of that disk under the bilinear map. Admitting one
series reactance as a second variable, the worst-case reflection over a scan
sector is shown to be quasiconvex, so the sector minimax is globally solvable
by bisection with a convex feasibility test. The results are applied to a
sixteen-element X-band array characterised by a finite-integration
time-domain solution with all sixteen ports excited in turn, and to the
corresponding $16\times24$ lattice. Every quantity reported is computed; no
measurement is included. For the array considered the isolated interior
element is matched to \SI{-16.94}{\decibel} while the broadside active
reflection is
$-6.6\pm\SI{1.7}{\decibel}$, the interval being propagated numerical
uncertainty and not a statistical dispersion. For the corresponding
two-dimensional lattice the criterion is violated at every scan angle
examined, and the minimax design improves the worst-case sector reflection
from $-2.6$ to \SI{-11.1}{\decibel}. Both matching designs, the relocated
probe for the line and the minimax for the lattice, are predictions of the
extracted circuit model: neither modified geometry has been re-solved in the
full-wave model, and they are reported as predictions throughout. The line and the lattice are two distinct
structures, the first solved as a finite sixteen-port problem and the second
as an infinite periodic one. Both are elements of an X-band radar aperture,
and the consequence is stated at the module: the aperture accepts $0.780$ of
the incident power at broadside and returns \SI{22.0}{\percent} of it, where
the isolated element returns \SI{2.0}{\percent}, so an acceptance criterion
written on the isolated reflection certifies a module load wrong by a factor
of $10.9$ in returned power. Consequences are confined to mismatch loss and
delivered power; no link budget, and no system-level radar analysis, is
attempted.
\end{abstract}


\section{Introduction}
\IEEEPARstart{T}{he} input impedance of an element in a scanning array is not
the impedance of that element in isolation. Under a progressive phase taper
the wave returning to a given port is the sum of its own reflection and every
contribution its neighbours couple back into it, and the resulting active
reflection coefficient differs from the isolated value in magnitude, in
frequency dependence, and in its variation with scan angle
\cite{amitay1972,pozar1994,hansen2009,mailloux2017}. The distinction is
established, but it is easily lost in a design flow that tunes a single
element and replicates it, because the isolated element can be made
arbitrarily well matched while the array is not matched at any scan angle in
the band.

The distinction is not academic for a radar front end. The active reflection
coefficient sets the power a transmit module actually delivers into the
aperture, the power it gets back, and the load its output stage drives, and on
a monostatic link the loss is paid twice. It is also scan dependent, so it is
not a constant that can be absorbed into a calibration factor: the load moves
with the beam. The array studied here is one row of the aperture of an X-band
radar under development, and Section~\ref{sec:radar} states the results at the
module in those terms --- accepted power, returned power and mismatch loss.
Quantities that need the rest of the link, detection range and range
resolution among them, are outside the scope of this paper and are not
claimed.

For probe-fed microstrip elements the usual remedy is to re-optimise the
probe position against a periodic model. That is sound but expensive: each
candidate position requires a fresh unit-cell solution, the search is
performed without any guarantee that a solution exists, and a failed search is
indistinguishable from a search that was not run long enough. The
contribution of this paper is to make both the existence question and the
optimisation tractable in closed form.

We begin from the observation that the impedance seen at a probe of a patch
in a periodic environment separates into a feed inductance, a transformer
ratio fixed by the probe position, and an array-loaded resonator that depends
on the lattice and the scan angle but not on the probe. Removing the first
two from a computed unit-cell impedance leaves the third, and the third is
what a probe-position search is searching against. One unit-cell sweep at an
arbitrary probe position therefore characterises the entire probe-position
design space.

Closed-form matching of a probe-fed patch has been treated for the isolated
element by Manteghi \cite{manteghi2009}, where the probe position and the
operating point are solved together from a cavity model. The treatment below
differs in three respects. The resonator is extracted from a computed
periodic solution rather than assumed, so it carries the array loading and
its variation with scan angle. The result is stated as a feasibility
inequality that settles existence before a solution is attempted. And the
sector-wide problem, which has no counterpart for an isolated element, is
shown to be convex in the sense required for global solution. Wide-angle
matching by loading the aperture \cite{hansen2009,mailloux2017} addresses the
same objective by a different mechanism and is complementary to what
follows.

The results are demonstrated on a sixteen-element X-band microstrip line
array, characterised by a full-wave transient solution in which every port is
excited in turn, and on the corresponding $16\times24$ lattice. The array is
a case in which the effect being described is large rather than marginal: the
isolated and active reflection coefficients differ by \SI{10.4}{\decibel},
and the sign of the scan loss is reversed with respect to the
projected-aperture law.

Table~\ref{tab:prior} places the present treatment against the closest
prior art. All quantities reported in this paper are computed. No prototype
was built and no experimental measurement is included;
Section~\ref{ssec:scope} states which computation each class of result comes
from, and Section~\ref{sec:conc} states the consequence for the strength of
the claims.

\begin{table*}[!t]
\caption{Relation to prior work}
\label{tab:prior}
\centering
\begin{tabular}{p{0.24\textwidth}p{0.22\textwidth}p{0.20\textwidth}p{0.26\textwidth}}
\toprule
& active-impedance design \cite{pozar1994,hansen2009,mailloux2017} & closed-form matching \cite{manteghi2009} & this work \\
\midrule
resonator model & implicit, in the solver & isolated element, cavity model & array-loaded, extracted from a periodic solution \\
scan dependence & retained & none & $\Zres(\theta_{s})$ per angle \\
existence decided before optimisation & no & no & \eqref{eq:feas} \\
reachable set characterised & no & no & disk, Thm.~\ref{thm:disk} \\
solver calls per candidate probe position & one & one & none after the first \\
sector-wide design & numerical search & not addressed & quasiconvex, globally solved \\
validation & varies & isolated element & finite 16-element array and $16\times24$ lattice \\
\bottomrule
\end{tabular}
\end{table*}

\emph{Contributions.} Individually, closed-form matching of an isolated
probe-fed patch \cite{manteghi2009}, Floquet extraction of array impedance
\cite{amitay1972,bhattacharyya2006} and convex methods in matching-network
synthesis \cite{boyd2004} are established. This paper contributes:

\begin{enumerate}
\item an extraction, \eqref{eq:extract}, giving an array-loaded resonator
      from \emph{one} unit-cell solution at an arbitrary probe position, so
      the whole probe-position design space costs one solver run rather than
      one per candidate (Section~\ref{ssec:decomp});
\item a proof that the impedances reachable by probe position and resonant
      length, at fixed frequency and scan angle, form a \emph{disk}
      (Theorem~\ref{thm:disk});
\item the necessary and sufficient feasibility condition
      $R_{p}\geq Z_{0}+X_{p}^{2}/Z_{0}$ that follows, with the matching point
      and, when it fails, the best attainable reflection, both in closed form
      (Proposition~\ref{prop:feas}, Corollary~\ref{cor:best});
\item a proof that the worst-case reflection over a scan sector is
      \emph{quasiconvex} in the probe ratio and one series reactance, hence
      globally solvable by bisection (Proposition~\ref{prop:qc});
\item validation on a finite sixteen-element line and on a $16\times24$
      periodic lattice, with an independent method-of-moments cross-check of
      the \emph{solution}, a decomposition of the residual disagreement, and a
      propagated uncertainty budget that includes the unit-cell adaptation
      residual (Sections~\ref{ssec:1d}--\ref{ssec:ucres});
\item a front-end statement of what the array-level mismatch costs at the
      module, in accepted and returned power rather than in pattern
      (Section~\ref{sec:radar}), and a demonstration that an acceptance
      criterion written on the isolated element understates the returned
      power by an order of magnitude.
\end{enumerate}

Both matching designs are \emph{predictions of the extracted circuit model}.
Neither modified geometry --- the relocated probe of Section~\ref{ssec:1d},
the probe and series capacitance of Section~\ref{ssec:2d} --- has been
re-solved in the full-wave model, and every number derived from them is
identified as a prediction where it appears. What is verified full wave is
the array as built and the extraction itself.

To the author's knowledge, items 2--4 have not been reported for an
array-loaded probe-fed element. Table~\ref{tab:prior} compares line by line.
Two supporting methodological results are also given: five consistency
identities for embedded-pattern post-processing (Section~\ref{ssec:ident}),
and a decomposition of a two-solver discrepancy into element-environment and
array-environment parts (Section~\ref{sec:cross}).

Throughout, $N=16$ is the number of elements, $d$ the element pitch,
$k=2\pi/\lambda_{0}$ the free-space wavenumber, $Z_{0}=\SI{50}{\ohm}$ the
reference impedance, and $\theta$ an angle measured from broadside in the
plane containing the array axis. The time convention is $e^{+j\omega t}$.

\section{Array, Numerical Model, and Validity Checks}
\label{sec:model}

\subsection{Element and lattice}
\label{ssec:geom}
\begin{figure*}[!t]
\centering
\includegraphics[width=\textwidth]{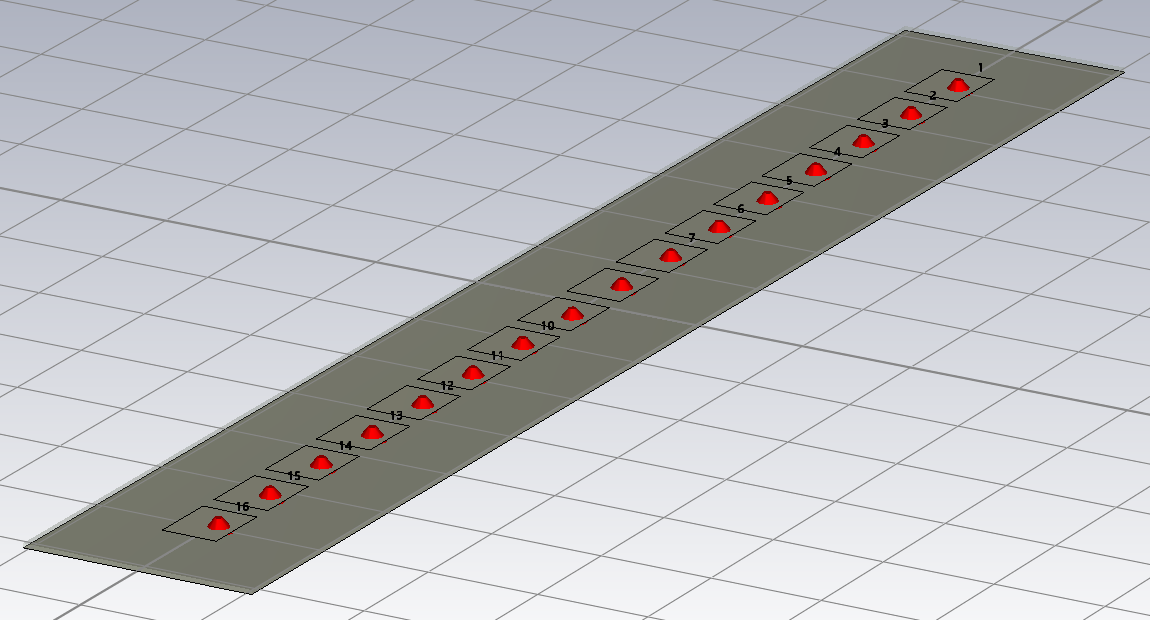}
\caption{The sixteen-element line as solved: probe-fed rectangular patches on
a common substrate over a finite ground plane, with the sixteen discrete
\SI{50}{\ohm} ports numbered as they are referred to throughout. Element~1 is
an edge element and element~8 an interior one; the numbering fixes the
scattering-matrix indices of Table~\ref{tab:coup} and the embedded patterns of
Table~\ref{tab:element}. The $16\times24$ lattice of Section~\ref{ssec:2d} is
a different model: one cell of it under periodic boundary conditions, never
solved as a finite structure.}
\label{fig:model}
\end{figure*}

Two structures appear in this paper and they are not the same object. The
first is a \emph{sixteen-element uniform linear array} of probe-fed
rectangular patches, Fig.~\ref{fig:model}, solved in full as a finite
sixteen-port problem with every port excited in turn; all scattering parameters, embedded element
patterns and synthesised array patterns below belong to it. The second is a
\emph{$16\times24$ planar lattice} of the same element on the same pitch,
represented by an infinite periodic unit cell under Floquet boundary
conditions and never solved as a finite structure; the active reflection
against scan angle in Section~\ref{sec:scan}, the extracted resonator
$\Zres(\theta_{s})$ and the matching design of Section~\ref{ssec:2d} belong to
it. The line is one row of that lattice in the sense that its element and
pitch are the same, and in no other sense: it has no neighbours in the E-plane
and the lattice model has no edges. Results are never transferred between them
without saying so, and Section~\ref{sec:scan} states what the line and the
cell do and do not corroborate in each other. Together they are the aperture
of an X-band radar under development, and every quantity reported in this
paper is a computed property of one of the two simulation models. 

The element is a rectangular patch on RT/duroid 5880
($\varepsilon_{r}=2.2$, $h=\SI{0.787}{\milli\metre}$,
$\tan\delta=0.0009$) at $f_{0}=\SI{9.35}{\giga\hertz}$, in the X band. The
transmission-line model gives a radiating-edge width
\begin{equation}
W=\frac{c}{2f_{0}}\sqrt{\frac{2}{\varepsilon_{r}+1}}=\SI{12.674}{\milli\metre},
\end{equation}
an effective permittivity $\varepsilon_{\mathrm{eff}}=2.054$ and, with the
Hammerstad fringing extension \cite{hammerstad1975},
$\Delta L=\SI{0.412}{\milli\metre}$ and a resonant length
$L=c/(2f_{0}\sqrt{\varepsilon_{\mathrm{eff}}})-2\Delta L
=\SI{10.363}{\milli\metre}$. The full-wave resonance occurs at
$L=\SI{9.968}{\milli\metre}$, \SI{3.8}{\percent} shorter; the width is
reproduced to four figures, so the closed-form fringing extension is
optimistic on this substrate and the cavity length is a starting value, not a
dimension.

The input resistance of the probe follows the standard cosine-squared law
\cite{carver1981,balanis2016},
\begin{equation}
R_{\mathrm{in}}(\ye)=R_{\mathrm{edge}}\cos^{2}\!\left(\frac{\pi \ye}{L}\right),
\label{eq:cos2}
\end{equation}
in which $\ye$ is measured \emph{from the radiating edge}, so that
$\ye=0$ gives the maximum resistance and $\ye=L/2$ at the patch centre gives
zero. The same argument defines the transformer ratio used throughout,
\begin{equation}
n^{2}(\ye)=\cos^{2}\!\left(\frac{\pi \ye}{L}\right),\qquad
n^{2}(0)=1,\quad n^{2}(L/2)=0 .
\label{eq:n2}
\end{equation}
A probe position may equally be quoted as an offset $y_{c}=L/2-\ye$ from the
patch centre. The two conventions are not interchangeable in
\eqref{eq:cos2}--\eqref{eq:n2} and every probe position below is given in
both. The as-built probe lies at $\ye=\SI{3.301}{\milli\metre}=0.3312L$,
equivalently $y_{c}=\SI{1.683}{\milli\metre}$, giving $n^{2}=0.256$.

Table~\ref{tab:geom} distinguishes the lengths that can be called the
aperture of the array. They differ by up to \SI{19}{\percent} and the
distinction matters in Section~\ref{sec:emb}, where a synthesised beamwidth is
compared with a theoretical one.

\begin{table}[!t]
\caption{Lattice and aperture dimensions}
\label{tab:geom}
\centering
\footnotesize
\begin{tabular}{lcr}
\toprule
quantity & symbol & value \\
\midrule
element pitch                    & $d$          & \SI{16.032}{\milli\metre}, $0.5000\lambda_{0}$ \\
end-element separation           & $(N-1)d$     & \SI{240.48}{\milli\metre} \\
periodic aperture                & $Nd$         & \SI{256.51}{\milli\metre} \\
conductor extent along axis      & $(N-1)d+W$   & \SI{253.15}{\milli\metre} \\
ground plane along the axis      & ---          & \SI{285.20}{\milli\metre} \\
ground plane across the axis     & ---          & \SI{42.03}{\milli\metre} \\
\bottomrule
\end{tabular}
\end{table}

The array axis is $x$ and the resonant dimension, hence the polarisation, is
$y$; the array therefore scans in its H-plane and the $\phi=0$ far-field cut
is the scan plane. Along the array axis the patch presents its width, leaving
a gap of \SI{3.36}{\milli\metre} between adjacent conductors.

\subsection{Solver configuration}
Two independent electromagnetic models are used and they are of different
kinds. The finite line, shown as modelled in, is solved
by the finite-integration technique in the time domain \cite{weiland1977} on
a structured hexahedral mesh, with all sixteen discrete ports excited in turn
so that the complete $16\times16$ scattering matrix and sixteen embedded
element patterns come from one set of runs; the run at each excitation is
terminated by a $\SI{-40}{\decibel}$ steady-state energy criterion rather
than at a fixed duration, and scattering parameters are obtained by Fourier
transform of the port signals. The infinite lattice is solved separately in
the frequency domain on an unstructured tetrahedral mesh with Floquet
boundaries, using adaptive mesh refinement at \SI{9.35}{\giga\hertz}, one
solution per scan angle. Table~\ref{tab:solver} lists every setting of both,
together with the extraction convention for each reported quantity, because
several results below depend on figures such as the reciprocity residual that
are meaningful only in conjunction with the discretisation that produced
them.

\begin{table}[!t]
\caption{Solver configuration. Every setting required to reproduce the three
numerical models.}
\label{tab:solver}
\centering
\footnotesize
\begin{tabular}{p{0.30\columnwidth}p{0.60\columnwidth}}
\toprule
\multicolumn{2}{l}{\textbf{Sixteen-element line} (finite integration, time domain \cite{weiland1977})}\\
\midrule
frequency range & \SIrange{8.8}{9.9}{\giga\hertz}, 1001 samples \\
boundaries & expanded open on all six faces; no symmetry planes \\
substrate & $\varepsilon_{r}=2.2$, $\mu_{r}=1$, $\tan\delta=0.0009$ specified at \SI{9.35}{\giga\hertz}, constant-$\tan\delta$ model, $\sigma=0$ \\
conductors & perfect electric conductor \\
ports & 16 discrete \SI{50}{\ohm} $S$-parameter ports, one per probe \\
mesh & hexahedral; 30 steps per wavelength near, 18 far; 40 steps per box near, 5 far; 803\,250 cells \\
excitation & all ports in turn; steady-state limit \SI{-40}{\decibel} \\
far-field monitors & 9.30, 9.35, \SI{9.40}{\giga\hertz}; $\theta$ and $\phi$ sampled at \SI{1}{\degree} \\
metal & zero-thickness PEC sheets; no finite conductivity, plating or roughness \\
probe & discrete port between patch and ground, via diameter not modelled \\
reference impedance & $Z_{0}=\SI{50}{\ohm}$ at every port, matching the intended feed network \\
time stepping & solver-selected from the Courant limit of the hexahedral mesh; run terminated by the \SI{-40}{\decibel} steady-state criterion, not by a fixed duration \\
active impedance & formed from the full $\mathbf{S}$ by \eqref{eq:gact} at each scan angle, then $\Zact=Z_{0}(1+\Gamma_{n})/(1-\Gamma_{n})$ \\
scan phase & $a_{m}=\exp(-jkx_{m}\sin\theta_{s})$ applied in post-processing to the solved $\mathbf{S}$, not re-solved per angle \\
\midrule
\multicolumn{2}{l}{\textbf{Floquet unit cell} (frequency domain, tetrahedral)}\\
\midrule
cell & one patch on $d\times d$ \\
frequency range & \SIrange{9.2}{9.5}{\giga\hertz} \\
boundaries & unit cell in $x$ and $y$ with constant scan angles; electric wall at $z_{\min}$; Floquet port at $z_{\max}$, two modes \\
mesh & tetrahedral, adaptive refinement \\
scan sweep & \SIrange{0}{60}{\degree} in \SI{10}{\degree} steps, seven separate solutions; $\phi_{s}=0$ \\
scan phase & imposed by the periodic boundary phase shift, one solution per angle \\
frequency samples & 61 across the band \\
convergence & adaptation terminated on the pass limit, residual $\approx0.08$ in $S$ \\
\midrule
\multicolumn{2}{l}{\textbf{Cross-check} (method of moments \cite{harrington1993})}\\
\midrule
element & same dimensions; ground plane \SI{42.03}{\milli\metre} along the resonant direction, $d$ along the array axis \\
mesh & default; 1644 triangles, 3972 tetrahedra, 7780 unknowns \\
\bottomrule
\end{tabular}
\end{table}

\subsection{Validity checks}
A passive structure satisfies $\mathbf{S}=\mathbf{S}^{\mathsf{T}}$. The
computed departure is $\max|S_{nm}-S_{mn}|=\num{8.9e-4}$ at
\SI{9.35}{\giga\hertz}, with an rms of \num{3.8e-4}, rising to \num{6.7e-3}
at the top of the swept band. The structure is also mirror-symmetric about
its centre, so with $\mathbf{J}$ the exchange matrix the solution must
satisfy $\mathbf{S}=\mathbf{J}\mathbf{S}\mathbf{J}$; the computed departure
is \num{4.0e-7}. Both residuals are propagated to the reported quantities in
Section~\ref{sec:unc}. Refining the mesh from 441\,216 to 803\,250 cells
moved the resonance by \SI{1.3}{\percent}; Section~\ref{sec:unc} states why
two levels do not constitute a convergence study in the sense of
\cite{roache1994}.

\subsection{Scope of the numerical evidence}
\label{ssec:scope}
Four distinct computations are reported and they are not interchangeable.
Scattering parameters and embedded element patterns of the finite line come
from the transient solution of Table~\ref{tab:solver}. Active reflection
against scan angle for the infinite lattice comes from the Floquet unit cell.
Unit-cell cross-check values and the single-element convergence calibration
come from the method-of-moments model. Everything in
Section~\ref{sec:design} beyond the extraction itself is a circuit
computation on impedances taken from the first two. Where the text below
refers to an impedance or a pattern as \emph{computed} or \emph{full-wave
extracted}, that is meant literally; the word \emph{measured} is not used of
any quantity in this paper, because no measurement was performed.

\section{Active Reflection of the Finite Array}
\label{sec:active}

At \SI{9.35}{\giga\hertz} the interior elements present $\SI{-16.94}{\decibel}$
and the edge elements $\SI{-14.13}{\decibel}$; the difference is the edge
condition, since an edge element has a neighbour on one side only. The band
over which all sixteen elements are simultaneously below
\SI{-10}{\decibel} is \SIrange{9.261}{9.413}{\giga\hertz}, a width of
\SI{152}{\mega\hertz}.

Nearest-neighbour coupling is $\SI{-12.75}{\decibel}$ between interior
elements and $\SI{-13.80}{\decibel}$ for the edge pair, with a mean of
$\SI{-12.98}{\decibel}$ over the fifteen adjacent pairs. The interior value
is the one relevant to the array. Couplings to more distant neighbours are
listed in Table~\ref{tab:coup}, together with the departure from reciprocity
of each entry. That departure is an absolute error of about
\num{9e-4} in $S$, so it is negligible for the adjacent pair and reaches
\SI{4.6}{\percent}, or \SI{0.39}{\decibel}, at the seventh neighbour; the
values quoted are the symmetrised $(\mathbf{S}+\mathbf{S}^{\mathsf{T}})/2$
and the last column states the precision to which each is determined.

\begin{table}[!t]
\caption{Coupling from element 1, and its numerical precision}
\label{tab:coup}
\centering
\begin{tabular}{lrrr}
\toprule
port pair & $\lvert S\rvert$ (dB) & $\arg S$ (deg) & precision (dB) \\
\midrule
adjacent, interior (8,9) & $-12.75$ & --- & 0.01 \\
adjacent, edge (1,2)     & $-13.80$ & $+135.4$ & 0.03 \\
second neighbour (1,3)   & $-22.41$ & $-42.4$  & 0.09 \\
third neighbour (1,4)    & $-27.36$ & $+127.8$ & 0.17 \\
fourth neighbour (1,5)   & $-29.84$ & $-54.7$  & 0.22 \\
seventh neighbour (1,8)  & $-36.25$ & $+136.6$ & 0.39 \\
\bottomrule
\end{tabular}
\end{table}

Two features matter. The phase alternates by approximately \SI{180}{\degree}
between successive neighbours, which is what a half-wavelength pitch
requires. And the decay with separation is slow, about \SI{5}{\decibel} per
element beyond the first, which is characteristic of a surface-wave
contribution rather than a purely reactive near field \cite{pozar1983}. A
calculation truncated at the first neighbour will not recover the total.

The single most consequential entries are the phases in the third column of
Table~\ref{tab:coup}: the self term and the nearest-neighbour term are in
phase to within \SI{5.3}{\degree}, and the remainder of this section shows
that this is the whole of the active-match problem.

Under a steering excitation $a_{m}=\exp(-jkx_{m}\sin\theta_{s})$ with
$x_{m}=(m-\tfrac{N+1}{2})d$, the active reflection coefficient and the array
mismatch efficiency are
\begin{align}
\Gamma_{n}(\theta_{s})&=\frac{1}{a_{n}}\sum_{m=1}^{N}S_{nm}a_{m},
\label{eq:gact}\\
\etamm(\theta_{s})&=1-\frac{\lVert\mathbf{S}\mathbf{a}\rVert^{2}}
{\lVert\mathbf{a}\rVert^{2}} .
\label{eq:etamm}
\end{align}
For uniform amplitude these satisfy
$\etamm=1-\langle|\Gamma_{n}|^{2}\rangle$, so the root-mean-square of
$|\Gamma_{n}|$ over the elements is the statistic consistent with
\eqref{eq:etamm} and is the one quoted here. The arithmetic mean differs by
\SI{0.04}{\decibel} and the two are not interchanged.

Reflection coefficients are quoted in decibels as
$20\log_{10}|\Gamma|$ throughout. At broadside,
\begin{equation}
\langle|\Gamma_{n}|^{2}\rangle^{1/2}=0.469\pm0.088,\qquad
\etamm=0.780\pm0.087,
\end{equation}
against an isolated interior-element reflection of $0.142$, that is
$-6.6\pm\SI{1.7}{\decibel}$ against \SI{-16.94}{\decibel}. The interval is
the propagated numerical uncertainty of Section~\ref{sec:unc}, dominated by
residual mesh error. It is not a dispersion across ports, a variation across
frequency, a standard deviation, or a statistical confidence interval; the
port-to-port spread of $|\Gamma_{n}|$ at broadside is separately $0.350$ to
$0.520$. The difference of \SI{10.4}{\decibel} between the isolated and
active values is an order of magnitude larger than the interval.

The mechanism is a phase coincidence. At broadside every $a_{m}$ is equal, so
\eqref{eq:gact} reduces to a row sum of $\mathbf{S}$, and
\begin{equation}
\arg S_{11}=\SI{130.1}{\degree},\qquad \arg S_{12}=\SI{135.4}{\degree}
\end{equation}
differ by only \SI{5.3}{\degree}, so the self term and the coupled terms add.
Accumulating the row of element~8 outward gives partial sums
$0.142$, $0.600$, $0.414$, $0.526$, $0.442$, $0.506$, $0.459$, $0.492$,
converging to $0.480$. Truncation at the first neighbour overestimates by
\SI{25}{\percent} and truncation at the second underestimates by
\SI{14}{\percent}.

Across scan angle the rms active reflection falls from $0.469$ at broadside
to $0.275$ at \SI{30}{\degree}, reaches a minimum near \SI{46}{\degree}, and
rises to $0.315$ at \SI{60}{\degree}. For the array considered and within the
sector examined, broadside is therefore the worst-matched angle. Nothing in
this observation is claimed to hold generally; it is a property of this
lattice, this substrate and this feed at this frequency, and the mechanism
identified above states the condition under which it arises.

Across frequency, driven at broadside, the rms active reflection is $-7.67$,
$-6.58$ and $\SI{-4.27}{\decibel}$ at 9.20, 9.35 and \SI{9.50}{\giga\hertz},
and does not reach $\SI{-10}{\decibel}$ anywhere between 9.0 and
\SI{9.7}{\giga\hertz}. The \SI{152}{\mega\hertz} quoted above is therefore the
bandwidth of the passive reflection, a manufacturing acceptance criterion, and
not an operating bandwidth of the aperture.

\section{Embedded Element Patterns and Array Synthesis}
\label{sec:emb}

\subsection{Superposition and de-embedding}
With $E_{n}(\theta)$ the embedded element pattern of port $n$, the field of
the fully excited array is the superposition
\begin{equation}
E(\theta)=\sum_{n=1}^{N}a_{n}E_{n}(\theta),
\label{eq:super}
\end{equation}
which is exact: mutual coupling and active mismatch are contained in the
computed $E_{n}$. It is not pattern multiplication, which replaces every
$E_{n}$ by one isolated pattern and contains no coupling.

Far-field phase is referenced to the global origin, so each $E_{n}$ carries a
position term and must be de-embedded,
\begin{equation}
\tilde{E}_{n}(\theta)=E_{n}(\theta)\exp(-j\sigma k x_{n}\sin\theta),\qquad
\sigma=\pm1 .
\label{eq:deemb}
\end{equation}
The sign is determined from the solver output rather than assumed. The phase of
$E_{1}/E_{8}$ must advance linearly in $\sin\theta$ with a slope set by the
known offset $x_{1}-x_{8}=-7d=\SI{-112.2}{\milli\metre}$; a least-squares fit
over $|\theta|\leq\SI{50}{\degree}$ returns \SI{-114.1}{\milli\metre}, which
fixes $\sigma=+1$.

\subsection{Consistency identities for far-field post-processing}
\label{ssec:ident}
Recombining exported far fields according to \eqref{eq:super} involves
conventions the solver does not enforce and no residual reports. Five
identities test them, each following from the physics rather than the code,
and each is stated with the discrepancy it detects on the present data. They
are inexpensive and we recommend them as routine.

\begin{enumerate}
\item[I1)] \emph{Continuity through the fold.} A principal-plane cut is
exported as two half-planes, at $\phi$ and $\phi+\pi$, whose reported phases
are referenced to local unit vectors with
$\hat{\boldsymbol{\phi}}(\phi+\pi)=-\hat{\boldsymbol{\phi}}(\phi)$ and
$\hat{\boldsymbol{\theta}}(\theta,\phi+\pi)=-\hat{\boldsymbol{\theta}}(-\theta,\phi)$.
The phase must run continuously through $\theta=0$: the step is
\SI{178.9}{\degree} without the correction and \SI{-1.1}{\degree} with it.
The omission is invisible in $|E|$, in any single-angle quantity, and in the
ratio of two patterns, since the sign is common to all elements at a given
angle; it becomes visible only where the fold is crossed, as when one element
is obtained by mirroring another.
\item[I2)] \emph{Slow residual phase.} A correctly de-embedded pattern varies
slowly: with $\sigma=+1$ in \eqref{eq:deemb} the residual phase of
$\tilde{E}_{1}-\tilde{E}_{8}$ varies by \SI{38}{\degree} across
$\pm\SI{60}{\degree}$, with $\sigma=-1$ by more than \SI{4000}{\degree}.
\item[I3)] \emph{On-axis agreement.} Both principal cuts describe the same
direction at $\theta=0$ and must agree there. The co-polar component is not
the same field component in both: with polarisation along $y$ it is $E_{\phi}$
in the $\phi=0$ cut and $E_{\theta}$ in the $\phi=\pi/2$ cut, where $E_{\phi}$
is \SI{44}{\decibel} down. With the correct components both report
\SI{5.035}{dBi}, agreeing in phase to \SI{0.001}{\degree}.
\item[I4)] \emph{Absolute, not normalised, cross-check.} A
pattern-multiplication script formed the array gain as $G_{e}+20\log_{10}N$
rather than $G_{e}+10\log_{10}N$, an error of \SI{12.04}{\decibel} that no
normalised comparison can reveal.
\item[I5)] \emph{Sphere integral.} A directivity pattern integrates to
$4\pi$. The two-cut $\cos^{2}\phi$ reconstruction used here integrates to
$1.16\times4\pi$, bounding the systematic error of any absolute directivity
derived from it at \SI{0.66}{\decibel}.
\end{enumerate}

A sixth check is numerical rather than physical: beamwidths must be obtained
by interpolating the $\SI{-3}{\decibel}$ crossings. Taking the outermost
samples above the threshold quantises the width to one grid step and biases it
low, returning \SI{6.00}{\degree} for a \SI{6.21}{\degree} beam on the
\SI{0.2}{\degree} grid used here.

\subsection{Radiometric conventions}
Directivity is normalised to radiated power, gain $G=\etarad D$ to accepted
power, and realised gain $G_{r}=(1-|\Gamma|^{2})G$ to incident power. All
element patterns quoted here are directivities. For element~8,
$\sum_{m}|S_{m8}|^{2}=0.158$, so \SI{15.8}{\percent} of the incident power
reaches the neighbouring loads; with a reported total efficiency of
\SI{81.4}{\percent} the inferred radiation efficiency is \SI{96.6}{\percent}.
Converting the array directivity to realised gain at broadside requires
$\etarad\etamm=0.754$, or $\SI{-1.23}{\decibel}$.

\subsection{Computed element patterns and synthesised array patterns}
\begin{figure*}[!t]
\centering
\includegraphics[width=\textwidth]{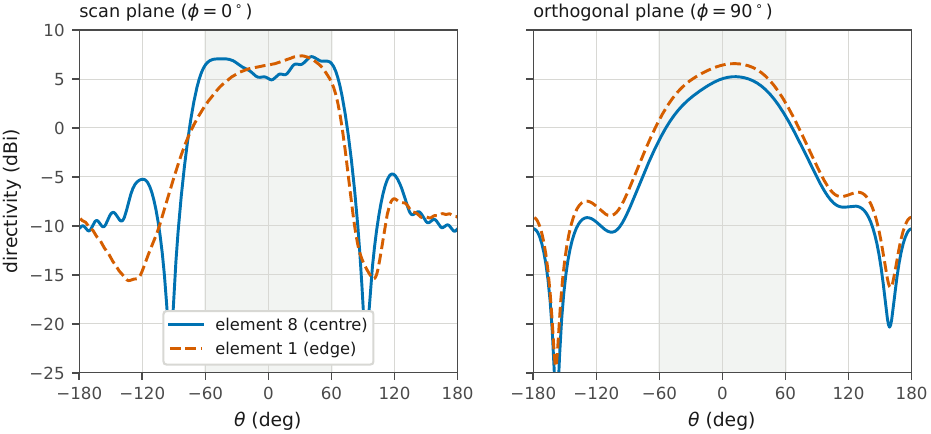}
\caption{Embedded element patterns of the interior and edge elements in the
scan plane and in the orthogonal plane. The shaded region is the intended
scan sector. The interior element is \SI{2.25}{\decibel} lower at broadside
than at its peak.}
\label{fig:element}
\end{figure*}

\begin{figure*}[!t]
\centering
\includegraphics[width=\textwidth]{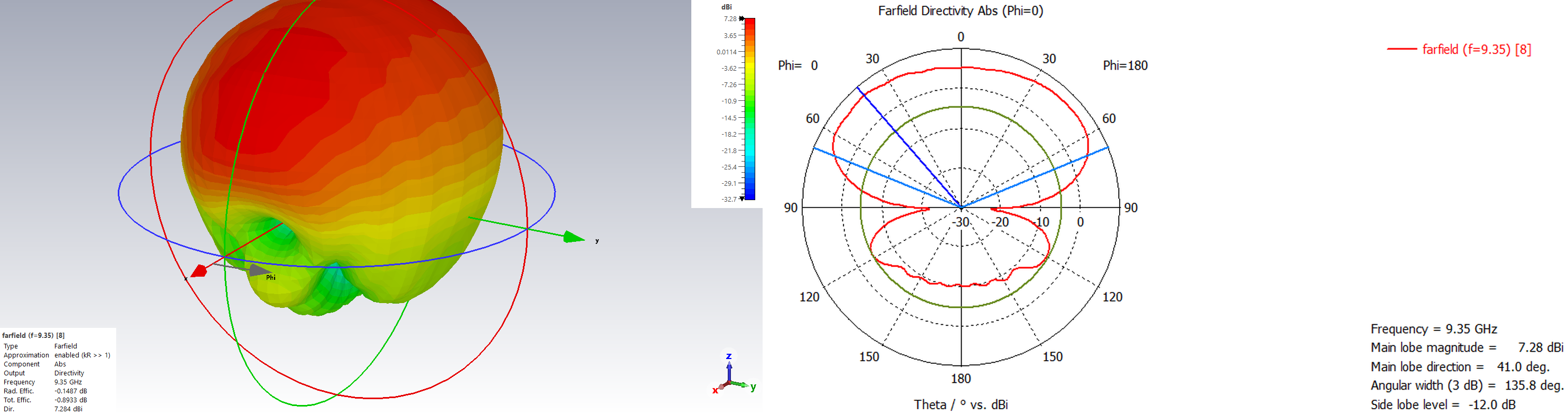}
\caption{The solver's own far-field readout for the interior element,
reproduced as exported. Left, the three-dimensional embedded directivity at
\SI{9.35}{\giga\hertz}; right, the scan-plane cut with the solver's own
computed parameters. The readout gives a main lobe of \SI{7.28}{dBi} at
\SI{41.0}{\degree} with a \SI{3}{\decibel} width of \SI{135.8}{\degree}, and
a total efficiency of \SI{-0.8933}{\decibel}, that is \SI{81.4}{\percent}.
Table~\ref{tab:element}, computed independently by the post-processing chain
of Section~\ref{ssec:ident} from the exported field files, gives
\SI{7.285}{dBi}, \SI{41.2}{\degree}, \SI{135.8}{\degree} and
\SI{81.4}{\percent}. The agreement is a check on the post-processing, not on
the solution: both come from the same solve.}
\label{fig:readout}
\end{figure*}

Fig.~\ref{fig:element} gives the computed embedded patterns and
Table~\ref{tab:element} their parameters. The interior element peaks
\SI{41.2}{\degree} off boresight and is \SI{2.25}{\decibel} lower on axis.
This is an independent statement of the result of Section~\ref{sec:active},
obtained from a different solver output. Fig.~\ref{fig:readout} reproduces the
solver's own readout for that element beside the pattern it exported: the two
agree on the peak, its angle and the beamwidth, which establishes that the
post-processing chain of Section~\ref{ssec:ident} reproduces what the solver
itself reports, and nothing more, since both rest on the same solution.

\begin{table}[!t]
\caption{Embedded element patterns at \SI{9.35}{\giga\hertz}}
\label{tab:element}
\centering
\begin{tabular}{lrr}
\toprule
quantity & element 8 & element 1 \\
\midrule
peak directivity                       & \SI{7.285}{\dBi} & \SI{7.375}{dBi} \\
angle of peak, scan plane              & \SI{41.2}{\degree} & \SI{31.6}{\degree} \\
directivity at broadside               & \SI{5.035}{dBi} & \SI{6.428}{dBi} \\
broadside deficit                      & \SI{2.25}{\decibel} & \SI{0.95}{\decibel} \\
\SI{3}{\decibel} width, scan plane     & \SI{135.8}{\degree} & \SI{105.6}{\degree} \\
\SI{3}{\decibel} width, orthogonal     & \SI{92.5}{\degree} & --- \\
accepted-power fraction                & 0.842 & 0.910 \\
total efficiency                       & \SI{81.4}{\percent} & \SI{87.9}{\percent} \\
\bottomrule
\end{tabular}
\end{table}

\begin{figure*}[!t]
\centering
\includegraphics[width=\textwidth]{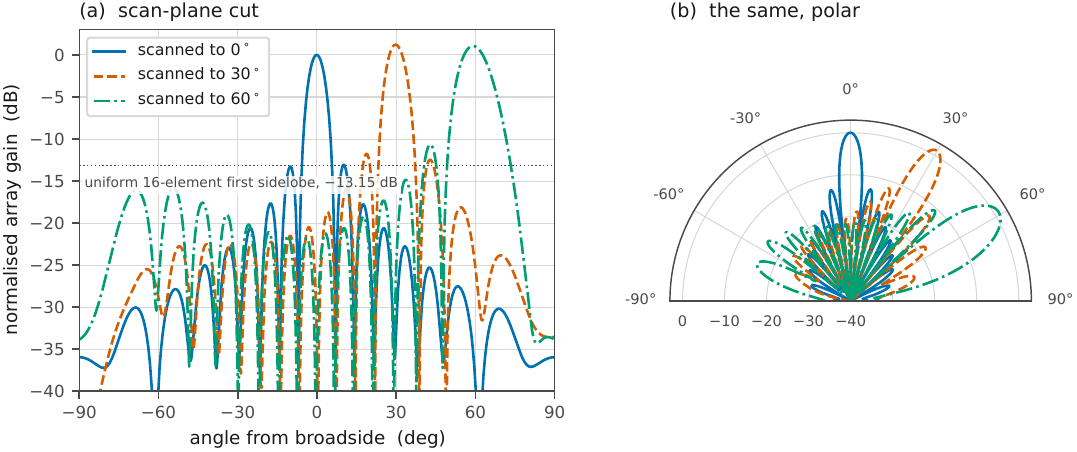}
\caption{(a) Array pattern in the scan plane, synthesised by superposition of
the computed embedded patterns and steered to 0, 30 and \SI{60}{\degree}. The
dotted line is the exact first-sidelobe level of a uniform sixteen-element
array, \SI{-13.147}{\decibel}. (b) The same three patterns in polar form over
the visible half-space.}
\label{fig:arraycut}
\end{figure*}

Equation~\eqref{eq:super} is evaluated with elements 2 to 15 represented by
element~8, element~1 by its own pattern, and element~16 by element~1 mirrored;
the influence of that assignment is quantified in Section~\ref{sec:unc}. The
reference for comparison is the exact array factor of $N$ isotropic sources,
evaluated numerically rather than through an asymptotic expression.
Table~\ref{tab:array} and Fig.~\ref{fig:arraycut} give the result.

\begin{table}[!t]
\caption{Synthesised array pattern against the exact array factor}
\label{tab:array}
\centering
\begin{tabular}{lrrr}
\toprule
& broadside & \SI{30}{\degree} & \SI{60}{\degree} \\
\midrule
synthesised peak, ref.\ broadside & \SI{0}{\decibel} & \SI{+1.22}{\decibel} & \SI{+1.06}{\decibel} \\
beam position                     & \SI{-0.1}{\degree} & \SI{29.8}{\degree} & \SI{59.0}{\degree} \\
synthesised \SI{3}{\decibel} width& \SI{6.214}{\degree} & \SI{7.443}{\degree} & \SI{11.797}{\degree} \\
exact array-factor width          & \SI{6.349}{\degree} & \SI{7.337}{\degree} & \SI{12.972}{\degree} \\
synthesised first sidelobe        & \SI{-13.07}{\decibel} & --- & --- \\
exact array-factor sidelobe       & \SI{-13.147}{\decibel} & --- & --- \\
\bottomrule
\end{tabular}
\end{table}

The exact array factor confirms the normalising length of
Table~\ref{tab:geom}: $0.886\lambda_{0}/(Nd)=\SI{6.346}{\degree}$ agrees with
the exact \SI{6.349}{\degree} to \SI{0.05}{\percent}, whereas $(N-1)d$ gives
\SI{6.769}{\degree} and is wrong by \SI{6.6}{\percent}. The synthesised beam is
\SI{2.1}{\percent} narrower than the exact array factor, consistent with the
mild inverse taper the embedded patterns impose. The sidelobe agreement is
not evidence: Section~\ref{sec:unc} shows the element-assignment choice moves
it over a \SI{0.90}{\decibel} range, an order of magnitude larger than the
\SI{0.08}{\decibel} discrepancy.

The scanned results are the substantive numerical result. A conventional
aperture loses gain as it scans, at least as fast as the projected-area law
$\cos\theta_{s}$, which costs \SI{3.01}{\decibel} at \SI{60}{\degree}. For the
array considered this synthesis gains $1.06\pm\SI{0.55}{\decibel}$, because
the embedded element pattern rises over most of the sector.

Broadside directivity is \SI{17.08}{dBi} from
$D_{\mathrm{array}}=D_{\mathrm{element}}+10\log_{10}N$ and \SI{16.89}{dBi}
from direct integration of the reconstructed three-dimensional pattern.
Identity~I5 bounds the systematic error of the latter at
\SI{0.66}{\decibel}, so the broadside directivity is
\SIrange{16.9}{17.1}{dBi} and no comparison at the tenth of a decibel is
meaningful.

\section{Scan Loss and the Periodic Limit}
\label{sec:scan}

For an infinite array the embedded element gain is related to the active
reflection coefficient by the element-pattern identity \cite{pozar1994},
\begin{equation}
G_{e}(\theta)=\frac{4\pi d_{x}d_{y}}{\lambda_{0}^{2}}\cos\theta
\left(1-|\Gamma(\theta)|^{2}\right)\etarad,
\label{eq:epi}
\end{equation}
whence the scan loss referenced to broadside,
\begin{equation}
\mathcal{L}(\theta)=-10\log_{10}
\frac{\cos\theta\left(1-|\Gamma(\theta)|^{2}\right)}{1-|\Gamma(0)|^{2}} .
\label{eq:scanloss}
\end{equation}
Equation~\eqref{eq:epi} holds in the periodic limit; for a finite line of
sixteen it is an approximation whose error is quantified in
Section~\ref{sec:cross}.

\begin{figure*}[!t]
\centering
\includegraphics[width=\textwidth]{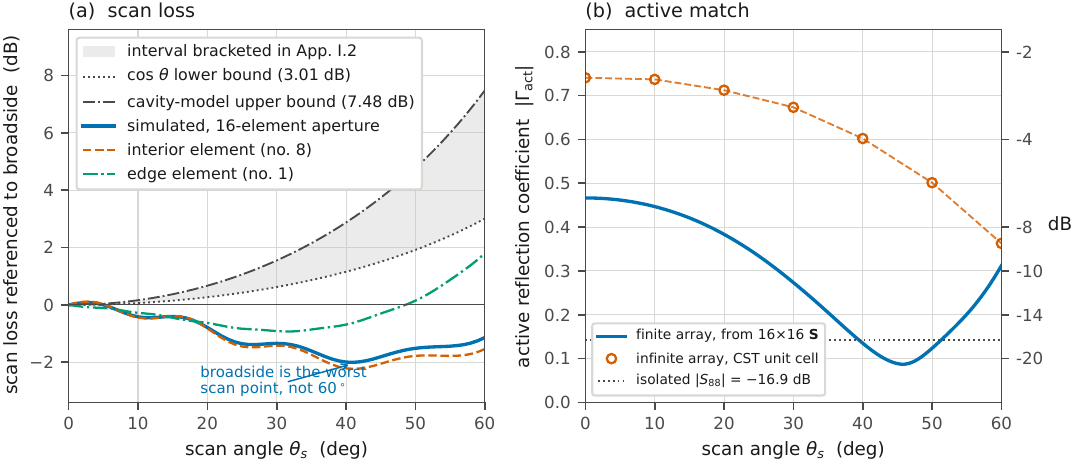}
\caption{(a) Scan loss of the sixteen-element line referenced to broadside,
against the two bounds a cavity model can offer: the projected-area law below
and an isolated-patch roll-off above. The computed curve falls outside both.
(b) Active reflection coefficient under a linear phase taper, from the
$16\times16$ scattering matrix and from the Floquet unit cell, against the
isolated-element value.}
\label{fig:scanloss}
\end{figure*}

Referenced to broadside, the sixteen-element line synthesised from its
embedded patterns gains \SI{1.14}{\decibel} at \SI{60}{\degree}, with the
best point of the sector at \SI{41}{\degree}, \SI{2.00}{\decibel} above
broadside, and a spread across the sector of \SI{2.09}{\decibel}
(Fig.~\ref{fig:scanloss}). A cavity model can bound the scan loss only
between the projected-area law below and an isolated-patch roll-off above, an
interval \SI{4.47}{\decibel} wide at the sector edge; both bounds presume an
element matched at broadside, and the computed curve falls outside both.

A Floquet unit cell of the same patch on the same lattice
\cite{bhattacharyya2006} gives an active reflection of $0.7405$ at broadside
falling monotonically to $0.3628$ at \SI{60}{\degree}, with no turning point.
The level is worse than the finite line because a two-dimensional lattice
couples in the E-plane as well, and E-plane coupling between patches at
half-wavelength spacing is the stronger of the two
\cite{pozar1982,schaubert1991}. These values carry the residual of about
$0.08$ noted in Table~\ref{tab:solver}.

What the three results establish is worth stating precisely. The scattering
matrix and the embedded patterns are two outputs of one solution on one mesh:
their agreement excludes post-processing error, since they travel through
different export files and different code, but says nothing about
discretisation error, which is common to both. The Floquet cell is a separate
solution with a different solver, mesh and boundary conditions, but of a
different structure, so it corroborates the mechanism and not the values. Only
the method of moments of Section~\ref{sec:cross} is an independent numerical
check, and only for the unit cell.

\begin{figure*}[!t]
\centering
\includegraphics[width=\textwidth]{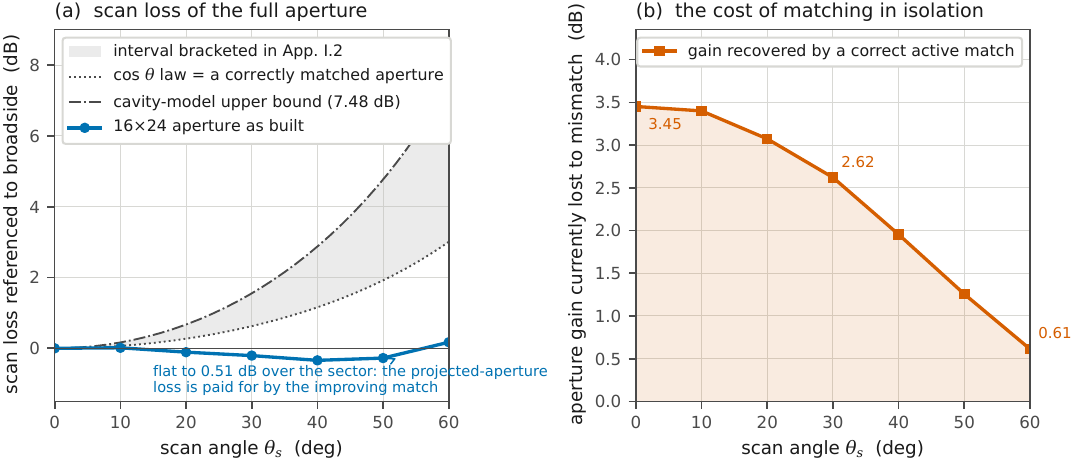}
\caption{The $16\times24$ aperture, from the Floquet unit cell. (a) Scan loss
referenced to broadside, against the interval a cavity model can bracket: the
computed curve is flat to \SI{0.51}{\decibel} across the sector because the
projected-aperture loss is paid for by an improving active match. (b) The
gain currently lost to mismatch at each angle, \SI{3.45}{\decibel} at
broadside falling to \SI{0.61}{\decibel} at \SI{60}{\degree}, which is what a
correct active match would recover. Both panels inherit the unit-cell
adaptation residual of Section~\ref{ssec:ucres}.}
\label{fig:aperture}
\end{figure*}

Substituting the unit-cell reflection into \eqref{eq:scanloss} gives the scan
loss of the $16\times24$ aperture for which the line is one row: $-0.11$,
$-0.34$ and $+\SI{0.17}{\decibel}$ at 20, 40 and \SI{60}{\degree}, an
total excursion of \SI{0.51}{\decibel}, Fig.~\ref{fig:aperture}(a). That flatness is purchased by
discarding $-10\log_{10}(1-|\Gamma(0)|^{2})=\SI{3.45}{\decibel}$ at broadside
in order to discard less at the sector edge. Of the 384 elements,
$14\times22=308$, or \SI{80.2}{\percent}, have a complete set of neighbours,
so the unit cell describes the interior rather than approximating it; a
direct solution would require of order $1.9\times10^{7}$ cells and 384
excitations.

\section{Impedance Decomposition and the Matching Design Space}
\label{sec:design}

\subsection{Decomposition}
\label{ssec:decomp}
Near resonance the impedance at the probe of a patch in a periodic
environment is a feed inductance in series with a resonator seen through the
transformer of \eqref{eq:n2},
\begin{equation}
\Zact=jX_{p}+n^{2}\Zres,\qquad X_{p}=\omega L_{p} .
\label{eq:decomp}
\end{equation}
Both $L_{p}$ and $\Zres$ are properties of the element and its lattice and
$n^{2}$ alone carries the probe position, so recovering $\Zres$ once yields
the whole design space without further electromagnetic solution. Given a
unit-cell impedance $Z^{(0)}$ computed at an arbitrary $n_{0}^{2}$,
\begin{equation}
\Zres(\theta_{s},f)=\frac{Z^{(0)}(\theta_{s},f)-j\omega L_{p}}{n_{0}^{2}} .
\label{eq:extract}
\end{equation}

\begin{figure*}[!t]
\centering
\includegraphics[width=\textwidth]{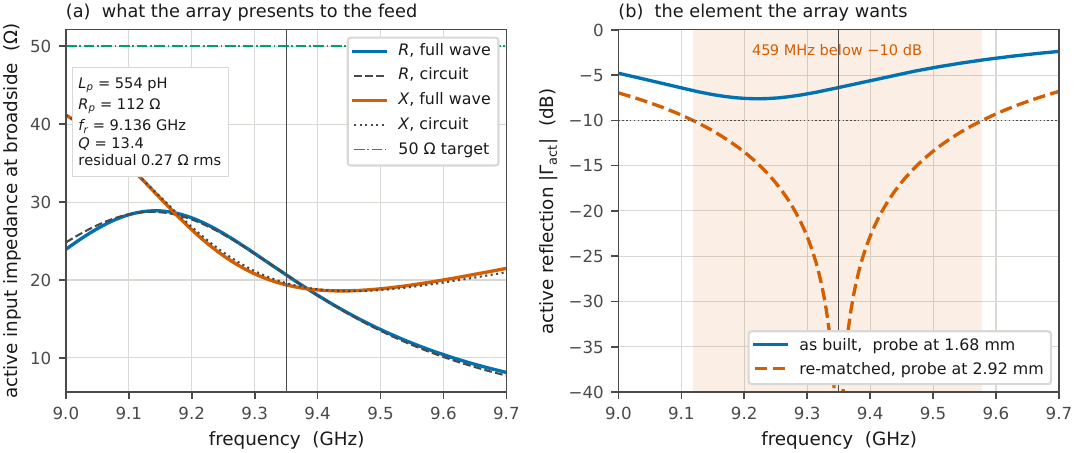}
\caption{(a) Active input impedance of the line at broadside, full wave
against the four-parameter model of \eqref{eq:decomp}; the residual is
\SI{0.27}{\ohm} rms over \SI{700}{\mega\hertz}. (b) Active reflection as
built and after the inversion of Section~\ref{ssec:feas}, which opens
\SI{459}{\mega\hertz} below \SI{-10}{\decibel} where the element as built has
none anywhere in the band.}
\label{fig:circuit}
\end{figure*}

Fitting \eqref{eq:decomp} with a single-pole resonator
$\Zres=R_{p}/(1+j\delta)$, $\delta=2Q(f-f_{r})/f_{r}$, to
\SI{700}{\mega\hertz} of the computed active impedance of the line at
broadside gives
\begin{equation}
L_{p}=\SI{553.7}{\pico\henry},\; R_{p}=\SI{112.3}{\ohm},\;
f_{r}=\SI{9.136}{\giga\hertz},\; Q=13.4
\end{equation}
with a residual of \SI{0.27}{\ohm} rms, so $X_{p}=\SI{32.53}{\ohm}$ at
\SI{9.35}{\giga\hertz} (Fig.~\ref{fig:circuit}); four parameters tracking a
full-wave impedance to a quarter of an ohm over a \SI{7.5}{\percent} band
indicates the topology is appropriate. For the two-dimensional lattice no pole
model is imposed and $\Zres$ is used directly from \eqref{eq:extract}.

\subsection{Feasibility of probe-only matching}
\label{ssec:feas}

The results of this subsection are stated as a theorem with explicit
hypotheses, because their strength depends entirely on the scope of those
hypotheses.

\begin{itemize}
\item[H1)] \emph{Topology.} At the port reference plane the active input
impedance is \eqref{eq:decomp}: a reactance $jX_{p}$ in series with a
resonator $\Zres$ seen through a real, frequency-independent transformer of
ratio $n^{2}$ set by the probe position through \eqref{eq:n2}.
\item[H2)] \emph{Resonator.} For Theorem~\ref{thm:disk} and
Proposition~\ref{prop:feas}, $\Zres(f)=R_{p}/(1+j\delta)$ with
$\delta=2Q(f-f_{r})/f_{r}$, $R_{p}>0$ and $0<Q<\infty$. $R_{p}$ is the
resonator resistance referred through a unit transformer, that is the value
of $\Rez\Zres$ at $\delta=0$.
\item[H3)] \emph{Loss.} $R_{p}>0$ strictly. It aggregates radiation loss,
dielectric loss and the power delivered to the neighbouring terminations; it
is not a conductor-loss term, so the perfect conductors of
Table~\ref{tab:solver} do not make it vanish. The hypothesis is not
cosmetic: Remark~\ref{rem:lossless} shows that the reachable set collapses to
a point as $R_{p}\to0$.
\item[H4)] \emph{Feed reactance.} $X_{p}=\omega L_{p}$ is evaluated at the
single frequency $f_{0}$ at which the whole construction is carried out, and
is taken independent of probe position over the range considered. The sign
convention is $Z=R+jX$ with $X>0$ inductive, and $X_{p}>0$.
\item[H5)] \emph{Length range.} The resonant length may be varied over an
interval that maps $\delta$, at the fixed $f_{0}$, onto a set containing the
value required. Remark~\ref{rem:range} quantifies this for the present
element.
\item[H6)] \emph{Reference.} $Z_{0}$ is real and positive; here
$Z_{0}=\SI{50}{\ohm}$, the impedance of the ports and of the intended feed
network.
\end{itemize}

The hypotheses are testable. $L_{p}$ comes from a wideband fit, but it can be
checked against the classical probe reactance
$X_{p}\simeq(\eta_{0}k_{0}h/2\pi)[\ln(2/k_{0}a)-\gamma]$ \cite{carver1981}:
inverting it for $X_{p}=\SI{32.53}{\ohm}$ at $h=\SI{0.787}{\milli\metre}$
gives a probe radius $a=\SI{0.170}{\milli\metre}$, an ordinary
\SI{0.34}{\milli\metre} via for this substrate. The extracted reactance is
therefore physical rather than a fitting artefact, which is the substantive
content of H1 and H4.

On identifiability: \eqref{eq:decomp} is unique \emph{given} $L_{p}$, since
for fixed $L_{p}$ the map $Z^{(0)}\mapsto\Zres$ of \eqref{eq:extract} is a
bijection; the pair $(L_{p},n_{0}^{2})$ is not separately identifiable from
one $Z^{(0)}$, because scaling $n_{0}^{2}$ and $R_{p}$ inversely leaves
$\Zact$ unchanged. The separation is fixed by taking $n_{0}^{2}$ from the
built geometry through \eqref{eq:n2} and $L_{p}$ from the fit, and the
preceding check is what makes that more than a convention; residual error in
it propagates linearly into $R_{p}$ and into the margin in \eqref{eq:feas}.

Finally, $\Zres$ is independent of the probe only within H1 and H4, at a
given scan angle, over the probe range for which $L_{p}$ is constant and over
the band of the fit. It is emphatically not independent of scan angle:
$\Zres(\theta_{s})$ varies by a factor of 2.1 in resistance across the sector,
and that variation is the array loading. Section~\ref{ssec:limits} bounds the
rest.

\begin{theorem}[Reachable impedance set]
\label{thm:disk}
Fix the frequency $f_{0}$ and the scan angle $\theta_{s}$, and let
$\Zact\in\mathbb{C}$ denote the active input impedance at the port reference
plane, at that frequency and that angle. Under H1--H5, the set of values
taken by $\Zact$ as the probe position varies over $n^{2}\in(0,1]$ and the
resonant length varies so that $\delta$ ranges over $\mathbb{R}$ is
\begin{equation}
\mathcal{R}=\mathcal{D}\setminus\{jX_{p}\},\quad
\mathcal{D}=\left\{Z:\left|Z-\left(\tfrac{R_{p}}{2}+jX_{p}\right)\right|
\leq\tfrac{R_{p}}{2}\right\},
\label{eq:disk1}
\end{equation}
whose closure is the closed disk $\mathcal{D}$ of centre
$R_{p}/2+jX_{p}$ and radius $R_{p}/2$. The boundary
$\partial\mathcal{D}\setminus\{jX_{p}\}$ is attained exactly at $n^{2}=1$,
and each interior point of $\mathcal{R}$ is attained at exactly one
$(n^{2},\delta)$.
\end{theorem}

\begin{remark}
\label{rem:variable}
$\mathcal{D}$ is a region of the complex plane of $\Zact$ at one frequency
and one scan angle. It is not a locus traced in frequency, and it is not the
reachable set at any other reference plane: moving the plane by a length of
transmission line maps $\mathcal{D}$ to another disk, and de-embedding a
shunt element does not preserve it at all. Every statement below is at
$f_{0}=\SI{9.35}{\giga\hertz}$ unless a frequency sweep is named explicitly.
\end{remark}

\begin{remark}
\label{rem:lossless}
H3 cannot be relaxed. The radius of $\mathcal{D}$ is $R_{p}/2$, so as
$R_{p}\to0$ the reachable set contracts to the single point $jX_{p}$ and no
match to any real $Z_{0}>0$ exists at any probe position or length. A
lossless resonator is unmatchable by a transformer, which is the circuit
statement of the fact that a real source impedance can only be matched to a
load that dissipates.
\end{remark}

\begin{remark}
\label{rem:range}
H5 is a restriction on the geometry, not a formality, because $\delta$ is
reached by changing the resonant length. For the present element
$f_{r}\propto L^{-1}$ to first order, so the length sweep of
Section~\ref{ssec:2d}, 8.5 to \SI{11.5}{\milli\metre}, maps to
$\delta\in[-3.4,+4.9]$ at $f_{0}$. The value required by
Proposition~\ref{prop:feas} is $\delta^{\star}=0.651$ against an as-built
$0.628$, so H5 holds with a margin of more than three units of $\delta$ on
either side and the corresponding length change is \SI{0.082}{\percent}. If
instead the length were constrained to a narrower interval
$[\delta_{\min},\delta_{\max}]$, the reachable set would be the corresponding
sub-family of arcs of \eqref{eq:circn} rather than the full disk, and
feasibility would additionally require $\delta^{\star}$ to lie in that
interval.
\end{remark}

\begin{proposition}[Feasibility]
\label{prop:feas}
Under H1--H6, an exact match $\Zact=Z_{0}$ is attainable if and only if
\begin{equation}
R_{p}\;\geq\;Z_{0}+\frac{X_{p}^{2}}{Z_{0}} ,
\label{eq:feas}
\end{equation}
in which case the solution is unique and given by
\begin{equation}
\delta^{\star}=\frac{X_{p}}{Z_{0}},\qquad
n^{2\star}=\frac{1}{R_{p}}\left(Z_{0}+\frac{X_{p}^{2}}{Z_{0}}\right).
\label{eq:feassol}
\end{equation}
\end{proposition}

Proposition~\ref{prop:feas} is the statement $Z_{0}\in\mathcal{D}$: since
$Z_{0}$ is real and positive it is distinct from $jX_{p}$, so membership of
$\mathcal{R}$ and of $\mathcal{D}$ coincide, and
$|Z_{0}-(R_{p}/2+jX_{p})|\leq R_{p}/2$ expands to
$Z_{0}^{2}-R_{p}Z_{0}+X_{p}^{2}\leq0$. Complete proofs of
Theorem~\ref{thm:disk} and Proposition~\ref{prop:feas} are given in
Appendix~\ref{app:a}.

\begin{corollary}[Best attainable when \eqref{eq:feas} fails]
\label{cor:best}
The bilinear map $\Gamma=(Z-Z_{0})/(Z+Z_{0})$ carries $\mathcal{D}$ to a disk
of centre $w_{0}$ and radius $r_{w}$ with
\begin{equation}
w_{0}=-\frac{\overline{B}}{A},\qquad
r_{w}=\frac{\sqrt{|B|^{2}-AC}}{|A|},
\label{eq:image}
\end{equation}
where, with $c=R_{p}/2+jX_{p}$ and $\rho=R_{p}/2$,
$A=|Z_{0}+c|^{2}-\rho^{2}$, $B=(Z_{0}+c)\overline{(Z_{0}-c)}+\rho^{2}$ and
$C=|Z_{0}-c|^{2}-\rho^{2}$. The smallest attainable reflection is therefore
\begin{equation}
|\Gamma|_{\min}=\max\left(0,\;|w_{0}|-r_{w}\right),
\label{eq:gmin}
\end{equation}
attained on $\partial\mathcal{D}$, that is at $n^{2}=1$. Condition
\eqref{eq:feas} is the case $|w_{0}|\leq r_{w}$.
\end{corollary}

Equation~\eqref{eq:gmin} converts an infeasible design into a quantitative
statement of how infeasible it is, without a search. For
$X_{p}=\SI{32.53}{\ohm}$ it gives $|\Gamma|_{\min}=0.561$, $0.280$ and
$0.085$ at $R_{p}=20$, $40$ and \SI{60}{\ohm}, and zero at and above
\SI{71.16}{\ohm}; a direct search over a $3000\times6000$ grid in
$(n^{2},\delta)$ reproduces each of these to six decimal places.

Two further consequences are worth stating.

\begin{remark}
The optimal detuning $\delta^{\star}$ depends only on the ratio of the feed
reactance to the reference impedance. It is independent of $R_{p}$ and of
$Q$, so the required offset of the resonance from the operating frequency is
fixed by the probe inductance alone and can be computed before the resonator
is characterised.
\end{remark}

\begin{remark}
Condition \eqref{eq:feas} is strictly stronger than the familiar requirement
$n^{2}R_{p}\geq2X_{p}$ for the reactance to be nullable, since
$Z_{0}+X_{p}^{2}/Z_{0}\geq2X_{p}$ by the arithmetic--geometric mean
inequality with equality only at $X_{p}=Z_{0}$. Nulling the reactance and
matching to $Z_{0}$ are different problems, and the weaker condition is
necessary but not sufficient. For the present element the two thresholds are
\SI{71.16}{\ohm} and \SI{65.06}{\ohm}, a difference of \SI{9.4}{\percent}.
\end{remark}

\begin{corollary}
\label{cor:general}
For a general (not single-pole) resonator at a fixed frequency and scan
angle, exact matching by probe position alone requires
$\Rez\Zres\geq Z_{0}$ and $\Imz\Zres\leq-X_{p}$ simultaneously, with
$n^{2}=Z_{0}/\Rez\Zres=-X_{p}/\Imz\Zres$. Two conditions on one variable are
generically inconsistent, so a probe alone is generically insufficient once
the resonance is not free to move.
\end{corollary}

\begin{figure*}[!t]
\centering
\includegraphics[width=\textwidth]{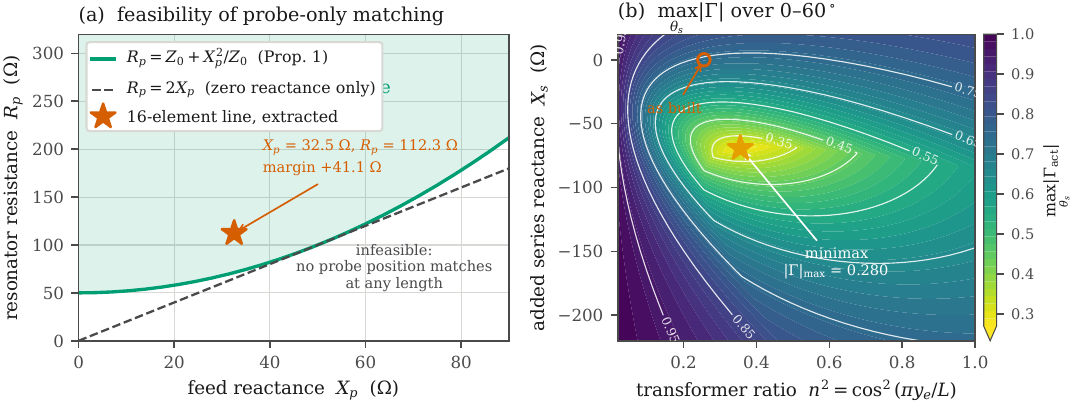}
\caption{(a) The feasibility boundary of Proposition~\ref{prop:feas} in the
$(X_{p},R_{p})$ plane, with the weaker zero-reactance bound and the
operating point of the sixteen-element line. (b) The sector objective
\eqref{eq:obj} for the two-dimensional lattice over the two design variables.
The sublevel sets are convex, as Proposition~\ref{prop:qc} requires, the
optimum is unique, and the active set at the optimum is the pair of sector
endpoints.}
\label{fig:design}
\end{figure*}

Fig.~\ref{fig:design}(a) draws \eqref{eq:feas} in the $(X_{p},R_{p})$ plane
together with the weaker bound and the operating point of the line,
which clears the threshold by \SI{41.1}{\ohm}.

\subsection{The sector minimax is quasiconvex}
\label{ssec:qc}
When \eqref{eq:feas} fails, a second degree of freedom is required. The least
intrusive is a series reactance $X_{s}$ at the feed, realisable as a gap in
the feed pad, giving
\begin{equation}
\Zin(\theta_{s};n^{2},X_{s})=n^{2}\Zres(\theta_{s})+j(X_{p}+X_{s}),
\label{eq:zin2}
\end{equation}
which is affine in the design vector $(n^{2},X_{s})$. The design objective
for a scanning aperture is the worst reflection over the intended sector,
\begin{equation}
F(n^{2},X_{s})=\max_{\theta_{s}\in\Theta}
\left|\frac{\Zin(\theta_{s};n^{2},X_{s})-Z_{0}}
{\Zin(\theta_{s};n^{2},X_{s})+Z_{0}}\right| .
\label{eq:obj}
\end{equation}

\begin{proposition}[Quasiconvexity]
\label{prop:qc}
For any finite $\Theta$, $F$ is quasiconvex on
$\{(n^{2},X_{s}):n^{2}>0\}$. Consequently \eqref{eq:obj} has no local minimum
that is not global, and its minimiser is obtained by bisection on
$r\in(0,1)$, each step testing feasibility of the convex quadratic system
\begin{multline}
(1-r^{2})\big(R_{\theta}^{2}+X_{\theta}^{2}\big)
-2Z_{0}(1+r^{2})R_{\theta}\\
+Z_{0}^{2}(1-r^{2})\leq0,\qquad \theta_{s}\in\Theta,
\label{eq:disk}
\end{multline}
with $R_{\theta}=n^{2}\Rez\Zres(\theta_{s})$ and
$X_{\theta}=n^{2}\Imz\Zres(\theta_{s})+X_{p}+X_{s}$.
\end{proposition}

The proof is in Appendix~\ref{app:b}. The mechanism is that constant-$|\Gamma|$
contours are circles in the impedance plane, so each sublevel set is the
preimage of a disk under an affine map, hence convex, and a finite
intersection of convex sets is convex.

\subsection{Application to the sixteen-element line}
\label{ssec:1d}
By Proposition~\ref{prop:feas}, $Z_{0}+X_{p}^{2}/Z_{0}=\SI{71.16}{\ohm}$
against $R_{p}=\SI{112.3}{\ohm}$, so \eqref{eq:feas} holds with a margin of
\SI{41.1}{\ohm} and no series element is needed.
Equation~\eqref{eq:feassol} gives $\delta^{\star}=0.6506$ and
$n^{2\star}=0.6337$, that is a probe at
$\ye=\SI{2.063}{\milli\metre}=0.207L$, equivalently
$y_{c}=\SI{2.921}{\milli\metre}$. The as-built detuning is $\delta=0.628$
against the required $\delta^{\star}=0.651$, so by Remark~\ref{rem:range} the
resonant length must increase by \SI{0.082}{\percent}, from 9.9679 to
\SI{9.976}{\milli\metre}, moving $f_{r}$ down by \SI{7.5}{\mega\hertz}. The
predicted active input impedance is $Z_{0}$ to machine precision and the
predicted active bandwidth below \SI{-10}{\decibel} is
\SI{459}{\mega\hertz}, where the as-built element has none anywhere in the
band. Both are outputs of the extracted circuit model of \eqref{eq:decomp} and
not of a full-wave solution of the relocated probe; Section~\ref{ssec:tol}
sets out what that leaves unaccounted for.

\subsection{Application to the two-dimensional lattice}
\label{ssec:2d}
For the lattice, $\Zres$ from \eqref{eq:extract} at \SI{9.35}{\giga\hertz} is
$86.65+j\SI{138.22}{\ohm}$ at broadside, $113.04+j\SI{130.60}{\ohm}$ at
\SI{30}{\degree} and $184.93+j\SI{20.51}{\ohm}$ at \SI{60}{\degree}. The
resistance condition of Corollary~\ref{cor:general} is met at every angle and
the reactance condition is violated at every angle: since
\begin{equation}
\Imz\Zin=X_{p}+n^{2}\Imz\Zres\;\geq\;X_{p}=+\SI{32.5}{\ohm}
\end{equation}
for all $n^{2}\in(0,1]$ whenever $\Imz\Zres>0$, no probe position matches
this lattice at any resonant length. Searching $n^{2}$ over its whole range
at broadside gives a best of $|\Gamma|=0.730$ against $0.7405$ as built, an
improvement of \SI{0.12}{\decibel}.

\begin{figure*}[!t]
\centering
\includegraphics[width=\textwidth]{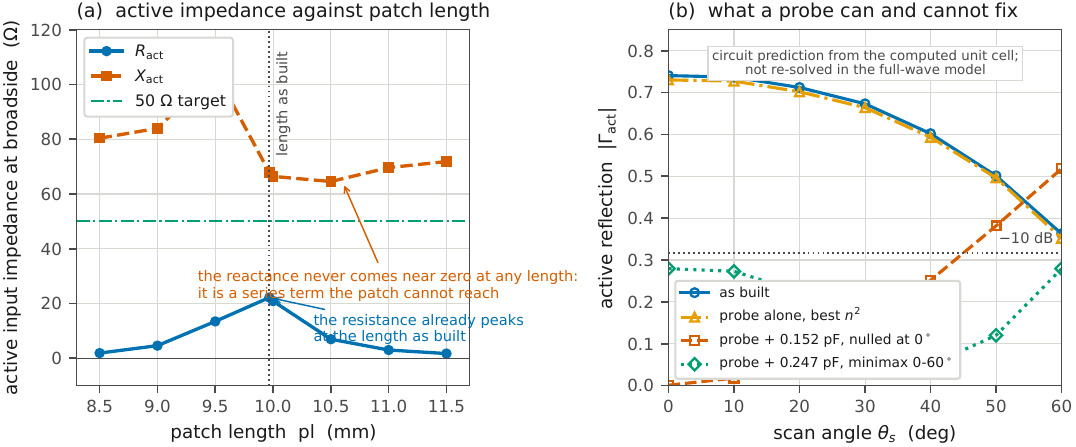}
\caption{(a) Full-wave active input impedance against patch length: the
resistance peaks at the length already in the model and the reactance remains
between $+65$ and $+\SI{105}{\ohm}$ throughout, never approaching zero, so the
lattice is not detuned. (b) Active reflection against scan angle for the
element as built, for the best probe position acting alone, for a design
nulled at broadside, and for the sector minimax of Section~\ref{ssec:qc}.}
\label{fig:lensweep}
\end{figure*}

This is a statement about the design space rather than about a particular
search, and it is confirmed independently by a full-wave sweep of eight patch
lengths from 8.5 to \SI{11.5}{\milli\metre}, Fig.~\ref{fig:lensweep}(a): the
active resistance peaks at the length already in the model and the active
reactance remains between $+65$ and $+\SI{105}{\ohm}$ throughout, never
approaching zero. The lattice is not detuned.

Proposition~\ref{prop:qc} is implemented as written: the objective
\eqref{eq:obj} is minimised over $\Theta=\{0,10,\dots,\SI{60}{\degree}\}$ by
bisection on the level $r$, each step testing feasibility of the intersection
of the seven disks \eqref{eq:disk} in the affine variables $(n^{2},X_{s})$,
which is a convex program; sixty bisection steps bracket the optimum to
\num{1e-18}. It returns $n^{2}=0.362$, that is
$\ye=\SI{2.934}{\milli\metre}=0.294L$, and $X_{s}=\SI{-69.0}{\ohm}$,
realisable as $C=\SI{0.247}{\pico\farad}$ at \SI{9.35}{\giga\hertz}, with the
worst sector reflection falling from $0.7405$ to $0.279$, that is from $-2.6$
to $\SI{-11.1}{\decibel}$, again as a prediction of the extracted circuit
model rather than a re-solved geometry. Sixteen local descents from different
starting points converge to that value to within \num{6e-17} and the
bisection agrees to the same precision, which is the behaviour
Proposition~\ref{prop:qc} predicts and would not be guaranteed without it. The active set is
$\{0,\SI{60}{\degree}\}$, the sector endpoints, the equioscillation signature
of a minimax solution.

A design that instead nulls the reflection at broadside alone gives
$n^{2}=0.577$ and $C=\SI{0.152}{\pico\farad}$ and degrades to
$\SI{-5.7}{\decibel}$ at \SI{60}{\degree}. Fig.~\ref{fig:lensweep}(b)
compares the two.

\subsection{Tolerance and limitations}
\label{ssec:tol}
The sensitivity of the broadside-nulled design to the series capacitance is
mild: a \SI{\pm10}{\percent} error leaves the reflection below
$\SI{-18}{\decibel}$ and a \SI{\pm20}{\percent} error below
$\SI{-11}{\decibel}$, with the \SI{-10}{\decibel} bandwidth exceeding
\SI{300}{\mega\hertz} in all cases, that being the full extent of the
unit-cell data.

The designs of this section are predictions of a circuit model built on
full-wave extracted impedance. The modified geometry has not been re-solved in
the full-wave model, and three effects are therefore unaccounted for. A gap
capacitance is not a pure series element: it adds shunt capacitance and has
its own self-resonance. Moving the probe perturbs the current distribution
and hence $R_{p}$, $f_{r}$ and $Q$, which the model holds fixed. And
\eqref{eq:extract} divides by the as-built $n_{0}^{2}$, so any error in the
probe-position convention propagates into $\Zres$. The predicted improvement
of \SI{8.5}{\decibel} in worst-case sector reflection is large with respect
to plausible values of all three, but it remains a prediction.

\subsection{Limits of validity}
\label{ssec:limits}
The hypotheses fail in identifiable ways, each with a symptom.

\emph{The probe itself is not modelled.} Both solvers drive the patch from a
discrete port between patch and ground with no via of finite diameter
(Table~\ref{tab:solver}), so the feed inductance $L_{p}$ that carries
$X_{p}=\SI{32.53}{\ohm}$, and with it the entire feasibility threshold
$Z_{0}+X_{p}^{2}/Z_{0}=\SI{71.16}{\ohm}$, is whatever the discrete-port
idealisation produces rather than the inductance of a physical via. This is
the most consequential idealisation in the paper for a probe-fed element: a
via of finite radius carries a different reactance, and the classical estimate
of Section~\ref{ssec:feas} is proportional to $\ln(2/k_{0}a)$, so a factor of
two in radius moves $X_{p}$ by roughly \SI{4}{\ohm} and the threshold by about
\SI{5}{\ohm}, a tenth of the \SI{41.1}{\ohm} margin the line enjoys. The
consistency check of Section~\ref{ssec:feas} is what keeps this honest: the
extracted $X_{p}$ inverts to a radius of \SI{0.170}{\milli\metre}, an ordinary
via on this substrate rather than an implausible one, so the idealisation is
at least in the right region. It is not a substitute for modelling the via,
and it is the first thing to change if the extracted $L_{p}$ ever disagrees
with a measurement. Nothing in Theorem~\ref{thm:disk} or
Propositions~\ref{prop:feas}--\ref{prop:qc} depends on where $X_{p}$ comes
from; the numerical thresholds do.

\emph{Large probe relocation.} H4 treats $L_{p}$ as position independent. The
designs here move the probe by $0.124L$ and $0.037L$; a move large enough to
alter the patch current distribution invalidates H1 and H4. Symptom: a fitted
$L_{p}$ that changes with probe position, testable with two solutions.

\emph{Change of dominant mode.} H2 presumes one resonance in the band. A
higher-order patch mode or a substrate mode entering the band voids
Theorem~\ref{thm:disk}. Symptom: the fit residual, here \SI{0.27}{\ohm} rms
over \SI{700}{\mega\hertz}.

\emph{Multiple resonances by design.} For stacked or aperture-coupled elements
the reachable set is the image of a two-pole locus and is not a
disk, so Theorem~\ref{thm:disk}, Proposition~\ref{prop:feas} and
Corollary~\ref{cor:best} lapse. Proposition~\ref{prop:qc} does not: its proof
uses only that $\Zin$ is affine in $(n^{2},X_{s})$ and never assumes a pole
model, so the sector minimax remains quasiconvex for any $\Zres(\theta_{s})$
whatever its modal content. This is the most portable result of the paper.

\emph{Scan blindness.} At a blindness angle $|\Gamma|\to1$ and $\Zres$ becomes
large and rapidly varying. The extraction remains valid pointwise and the
minimax is then dominated by that angle, returning a value bounded away from
zero, which is the correct answer rather than a failure. None appears in the
sector examined: the unit-cell reflection is monotone from 0 to
\SI{60}{\degree}.

\emph{Very strong or probe-dependent coupling.} All coupling is absorbed into
$\Zres(\theta_{s})$, which is exact at a given angle by construction; what
the decomposition cannot represent is coupling that depends on probe
position. For patches, coupling is dominated by the radiating edges and by
surface waves rather than by the feed, so the dependence is expected to be
weak, but this is an assumption.

\emph{Large frequency excursions, dispersion and nonlinearity.} H2 and H4
hold $L_{p}$, $R_{p}$ and $Q$ constant over the fitting band, and the
constant-$\tan\delta$ material of Table~\ref{tab:solver} carries no
dispersion. Strongly dispersive substrates, or a wider band, require the
extraction to be repeated per band, which is inexpensive. The whole
construction is linear: it does not apply to elements with ferrite, varactor
or other nonlinear loading, for which no single $\Zres$ exists.

\section{Cross-Validation and Attribution}
\label{sec:cross}

An independent model was built using a method of moments on a
surface--volume tetrahedral mesh \cite{harrington1993}. The two models are
not the same structure: the second solves one element rather than sixteen,
forms the array pattern by pattern multiplication and therefore contains no
mutual coupling, and places its element on a ground plane truncated to
$0.5\lambda_{0}$ along the array axis rather than $8.9\lambda_{0}$. The last
difference lies in the plane in which the two are compared.

\emph{How independent, and independent of what.} The two models are
independent as to \emph{solution} error and identical as to \emph{model}
error, and the distinction decides what the cross-check is worth. They are
given the same nominal geometry of Table~\ref{tab:geom} and the same material
constants of Table~\ref{tab:solver}: the same dimensions, the same
$\varepsilon_{r}$ and $\tan\delta$, the same perfect zero-thickness
conductors, the same idealised probe with no via diameter. In that sense the
moment method does not validate the model; it recomputes the same idealised
object by another route. What it does not share is everything between that
object and the numbers: integral against differential formulation;
unbounded domain with a free-space Green's function against a truncated
volume with an absorbing boundary, so the moment method has no mesh
truncation error in the exterior at all; unstructured tetrahedra against
structured hexahedra; direct factorisation of a dense matrix against explicit
time stepping; and separate implementations. A discretisation error giving
the same wrong answer in both would need two unrelated mechanisms, and
agreement between them is therefore evidence that the discretisation, the
meshing and the post-processing are not where an error lives. It is not
evidence about the object. A wrong permittivity, the unmodelled via
inductance of Section~\ref{ssec:limits}, a conductor thickness that matters,
a fabrication tolerance: each of these moves both solvers by the same amount
and the comparison reports perfect agreement. That class of error is what a
measurement addresses and what no second solver can. The claim made here is
the narrow one: the numbers reported are the numbers this model implies, to
the precision stated.

\emph{Agreement, and its size against the uncertainty budget.} On the unit
cell, where the structures are the same, the active reflection is $0.7405$,
$0.6731$ and $0.3628$ against $0.5971$, $0.5550$ and $0.3572$ at $0$, $30$
and \SI{60}{\degree}. At \SI{60}{\degree} they differ by $0.0056$, which is
\SI{1.5}{\percent} and is smaller than the $0.083$ uncertainty assigned to
that quantity in Table~\ref{tab:unc}; the two methods therefore agree to
within the numerical uncertainty of either. At broadside they differ by
$0.14$, which is \SI{19}{\percent} and is $1.6$ times the assigned
uncertainty of $0.088$, so the broadside disagreement is real and is not
explained by discretisation alone. Its most likely origin is the
adaptive-refinement residual of about $0.08$ carried by the
finite-integration unit cell (Table~\ref{tab:solver}), which is of the right
size.

\begin{figure*}[!t]
\centering
\includegraphics[width=\textwidth]{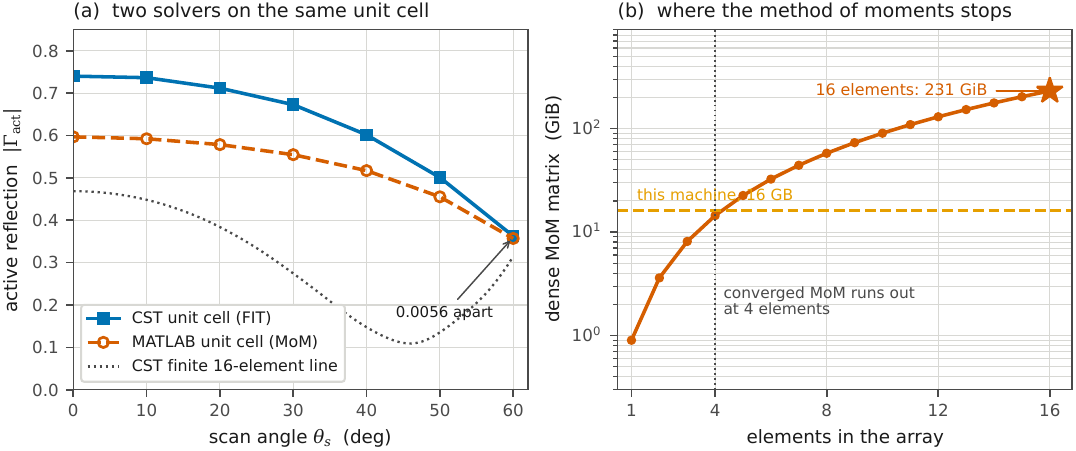}
\caption{(a) The two solvers on the same unit cell, with the finite
sixteen-element line shown for contrast: the cell is the structure the two
methods share, and they close to $0.0056$ at \SI{60}{\degree} of scan. The
finite line is a different structure and is not expected to coincide with
either. (b) Why the finite line was not solved by the method of moments: the
dense matrix grows as the square of the unknown count, so sixteen elements
require \SI{231}{\gibi\byte} and the largest converged array that fits the
\SI{16}{\giga\byte} available is four elements.}
\label{fig:mom}
\end{figure*}

\emph{Computational limit.} The finite line is beyond the method of moments
on the hardware used: a converged element requires 7780 unknowns, so sixteen
require 124\,480 and a dense complex matrix of \SI{231}{\gibi\byte}. Memory
scales as $16N_{\mathrm{unk}}^{2}$ bytes and factorisation as
$N_{\mathrm{unk}}^{3}$, so the largest converged array fitting in
\SI{16}{\giga\byte} is four elements, Fig.~\ref{fig:mom}(b), and every
manually coarsened mesh
returned an element impedance 25 to \SI{104}{\percent} from the converged
value. One coarsening did not fail visibly: raising the minimum edge length
to \SI{2.5}{\milli\metre}, above the \SI{0.787}{\milli\metre} substrate,
allowed a sixteen-element solve to complete with the dielectric unresolved
and the patch not resonating, giving $|S_{11}|=\SI{-2.2}{\decibel}$ against
$\SI{-12.6}{\decibel}$ isolated. The guard is to re-solve one element on the
reduced mesh against a converged reference before accepting the array.

\begin{figure*}[!t]
\centering
\includegraphics[width=\textwidth]{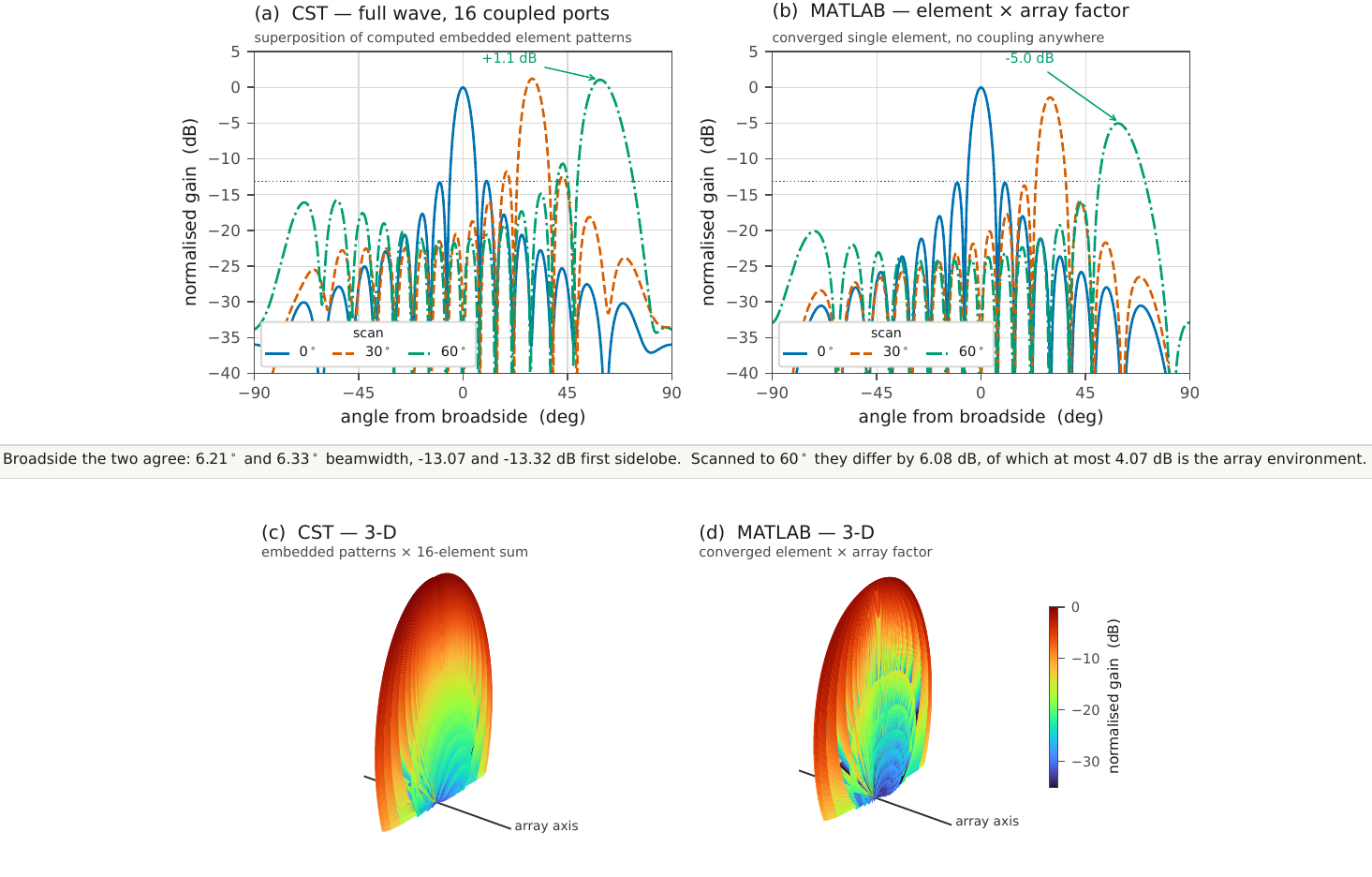}
\caption{Array pattern in the scan plane from both methods, and the same
patterns in three dimensions. (a), (c) superposition of sixteen computed
embedded element patterns, containing mutual coupling and active mismatch;
(b), (d) one converged element multiplied by the array factor, containing
neither. The two agree at broadside and diverge by \SI{6.08}{\decibel} at
\SI{60}{\degree} of scan; Table~\ref{tab:attr} decomposes the divergence.}
\label{fig:compare}
\end{figure*}

\emph{Array patterns.} Table~\ref{tab:cross} and Fig.~\ref{fig:compare}
compare the two. The broadside directivity agreement of
\SI{0.01}{\decibel} is fortuitous and is not a validation, since identity~I5
bounds the systematic error of the first figure at \SI{0.66}{\decibel}; what
the broadside row establishes, at the precision it supports, is that
geometry, units, element model and array factor are consistent in both to
better than a decibel. The beamwidth agreement of \SI{0.11}{\degree}, against
an element-assignment spread of \SI{0.16}{\degree}, carries the same weight.

\begin{table}[!t]
\caption{Array patterns from the two methods}
\label{tab:cross}
\centering
\begin{tabular}{lrrr}
\toprule
& finite integration & moment method & difference \\
\midrule
broadside directivity   & \SI{16.89}{dBi} & \SI{16.88}{dBi} & \SI{0.01}{\decibel} \\
broadside \SI{3}{\decibel} width & \SI{6.214}{\degree} & \SI{6.327}{\degree} & \SI{0.11}{\degree} \\
first sidelobe          & \SI{-13.07}{\decibel} & \SI{-13.32}{\decibel} & \SI{0.25}{\decibel} \\
scanned \SI{30}{\degree}& \SI{+1.22}{\decibel} & \SI{-1.40}{\decibel} & \SI{2.62}{\decibel} \\
scanned \SI{60}{\degree}& \SI{+1.06}{\decibel} & \SI{-5.02}{\decibel} & \SI{6.08}{\decibel} \\
\bottomrule
\end{tabular}
\end{table}

\emph{Attribution.} Attributing the whole \SI{6.08}{\decibel} at
\SI{60}{\degree} to mutual coupling is wrong by about \SI{2}{\decibel}. In
pattern multiplication the peak amplitude of the array factor is invariant
under scan, so the entire scanned variation of that curve is its element
pattern. Referring both to the projected-area law gives Table~\ref{tab:attr},
which separates what belongs to the array from what belongs to the element
model.

\begin{table}[!t]
\caption{Decomposition of the divergence at \SI{60}{\degree} of scan}
\label{tab:attr}
\centering
\begin{tabular}{lr}
\toprule
contribution & value \\
\midrule
projected-area law, $-10\log_{10}\cos\SI{60}{\degree}$ & \SI{-3.01}{\decibel} \\
moment-method element, computed roll-off               & \SI{-5.09}{\decibel} \\
\quad excess: element environment                      & \SI{-2.08}{\decibel} \\
synthesised array, computed                            & \SI{+1.06}{\decibel} \\
\quad excess: array environment                        & \SI{+4.07}{\decibel} \\
\quad\quad accounted for by \eqref{eq:epi}             & \SI{+0.62}{\decibel} \\
\quad\quad unaccounted for                             & \SI{+3.45}{\decibel} \\
\bottomrule
\end{tabular}
\end{table}

At most \SI{4.07}{\decibel} of the divergence is attributable to placing the
element in the array; the remaining \SI{2.08}{\decibel} is a property of the
isolated element of the second model, whose ground plane is truncated to half
a wavelength along the array axis and which is therefore more directive than
$\cos\theta$. A larger ground plane in the scan plane would make it more
directive still, so \SI{4.07}{\decibel} is a lower bound on the
array-environment term. Both figures are large against the
\SI{0.55}{\decibel} uncertainty that Table~\ref{tab:unc} assigns to the
\SI{60}{\degree} peak. Of the \SI{4.07}{\decibel}, the periodic identity
\eqref{eq:epi} evaluated with the finite-array reflection coefficients
accounts for \SI{0.62}{\decibel}; the remaining \SI{3.45}{\decibel} is the
embedded pattern of a finite array, which \eqref{eq:epi} does not govern.

\section{Numerical Uncertainty}
\label{sec:unc}

\subsection{Sources and budget}
Each source below is a computed solver residual propagated through the same
post-processing chain that produces the reported quantity.

Enforcing $\mathbf{S}=(\mathbf{S}+\mathbf{S}^{\mathsf{T}})/2$ changes the
broadside active reflection by less than \num{1e-4} and the
\SI{60}{\degree} value by \num{4e-4}; enforcing mirror symmetry changes every
quantity by less than \num{1e-4}. Decimating the exported far field from
\SI{1}{\degree} to \SI{2}{\degree} changes the beamwidth by
\SI{0.005}{\degree} and the \SI{60}{\degree} scanned peak by
\SI{0.017}{\decibel}, and coarsening the interpolation grid from
\SI{0.02}{\degree} to \SI{0.10}{\degree} changes nothing at the third
decimal. The two dominant terms are therefore the operating point, used as a
surrogate for residual mesh error, and the assignment of embedded patterns to
the twelve elements that were not exported. Four defensible assignments give
beamwidths from 6.214 to \SI{6.375}{\degree}, first sidelobes from $-12.169$
to $\SI{-13.069}{\decibel}$, and \SI{60}{\degree} peaks from $+1.023$ to
$\SI{+1.570}{\decibel}$. Table~\ref{tab:unc} collects the budget.

\begin{table}[!t]
\caption{Uncertainty budget}
\label{tab:unc}
\centering
\begin{tabular}{lrr}
\toprule
quantity & value & uncertainty \\
\midrule
$\langle|\Gamma|^{2}\rangle^{1/2}$, broadside & 0.4688 & 0.088 \\
$\langle|\Gamma|^{2}\rangle^{1/2}$, \SI{30}{\degree} & 0.2746 & 0.070 \\
$\langle|\Gamma|^{2}\rangle^{1/2}$, \SI{60}{\degree} & 0.3146 & 0.083 \\
mismatch efficiency, broadside & 0.7803 & 0.087 \\
\SI{3}{\decibel} beamwidth & \SI{6.214}{\degree} & \SI{0.161}{\degree} \\
first sidelobe & \SI{-13.07}{\decibel} & \SI{0.90}{\decibel} \\
peak at \SI{30}{\degree} scan & \SI{+1.220}{\decibel} & \SI{0.20}{\decibel} \\
peak at \SI{60}{\degree} scan & \SI{+1.064}{\decibel} & \SI{0.55}{\decibel} \\
absolute directivity (systematic) & \SI{16.9}{dBi} & \SI{0.7}{\decibel} \\
\midrule
\multicolumn{3}{l}{\emph{unit cell, from the \num{0.08} adaptation residual}}\\
$\Rez\Zres$, broadside & \SI{86.65}{\ohm} & $^{+32.7}_{-29.1}$\,\si{\ohm} \\
$\Imz\Zres$, \SI{60}{\degree} & \SI{20.51}{\ohm} & $^{+35.3}_{-33.3}$\,\si{\ohm} \\
minimax worst-case $|\Gamma|$ & 0.279 & $^{+0.051}_{-0.031}$ \\
\bottomrule
\end{tabular}
\end{table}

No term above estimates the mesh discretisation error of the sixteen-element
solution. Two mesh levels were run, at 441\,216 and 803\,250 cells, a ratio
of 1.82 in cell count and therefore 1.22 in linear cell size. Richardson
extrapolation from two levels requires an assumed order, and at a refinement
ratio of 1.22 the extrapolation is ill-conditioned: for second-order
convergence the remaining error is approximately twice the observed change
between levels rather than a fraction of it \cite{roache1994}. Treating the
observed \SI{1.3}{\percent} shift as the residual is therefore optimistic,
and the surrogate term, already dominant, is the one most likely to be
understated. A three-level study at a ratio near two in linear cell size,
which for this model means of order $6\times10^{6}$ cells, was not performed.
The conclusions below are stated in a form that survives this omission: the
differences on which they depend are 10.4 and \SI{4.07}{\decibel}, against
uncertainties of 1.7 and \SI{0.55}{\decibel}.

\subsection{The unit-cell residual, and what rests on it}
\label{ssec:ucres}
The frequency-domain unit cell terminated its adaptive refinement on the pass
limit rather than on a convergence target, at a residual of about $0.08$ in
$S$ (Table~\ref{tab:solver}). That is not a small number against the
quantities it feeds, and two conclusions rest on it: the non-existence result
for the lattice, Section~\ref{ssec:2d}, and the sector minimax design. Both
are therefore tested against it rather than asserted through it.

The residual is treated as an uncertainty of magnitude $0.08$ on the
unit-cell reflection coefficient, of unknown phase, and carried through
\eqref{eq:extract} for every phase on the circle. The extracted resonator
moves substantially: at broadside $\Rez\Zres$ spans 57.6 to \SI{119.3}{\ohm}
about its nominal \SI{86.65}{\ohm}, and at \SI{60}{\degree} $\Imz\Zres$ spans
$-12.8$ to $+\SI{55.9}{\ohm}$ about its nominal $+\SI{20.5}{\ohm}$. The
extracted values are therefore not precise, and no argument in this paper
should depend on their precision.

The non-existence conclusion does not. As written in Section~\ref{ssec:2d} it
invokes $\Imz\Zres>0$, which the perturbation can violate at
\SI{60}{\degree}, so it is restated here in the form the data support. Since
$\Imz\Zin=X_{p}+n^{2}\Imz\Zres$ with $n^{2}\in(0,1]$, the minimum over probe
position is $X_{p}$ when $\Imz\Zres\geq0$ and $X_{p}+\Imz\Zres$ when
$\Imz\Zres<0$, so
\begin{equation}
\begin{split}
&\Imz\Zres>-X_{p}\ \text{at every angle}\\
&\qquad\Longrightarrow\ \Imz\Zin>0\ \text{for every probe position} .
\end{split}
\label{eq:robust2d}
\end{equation}
Over the whole perturbation and all seven angles the worst extracted
reactance is $\SI{-12.8}{\ohm}$, against $-X_{p}=\SI{-32.5}{\ohm}$, so
\eqref{eq:robust2d} holds with \SI{19.8}{\ohm} of margin and
$\Imz\Zin\geq\SI{19.8}{\ohm}$ throughout. No probe position matches this
lattice at any angle in the sector, for any perturbation of the unit-cell
reflection of magnitude $0.08$. The resistance condition of
Corollary~\ref{cor:general} is likewise satisfied at every angle for every
perturbation, the smallest $\Rez\Zres$ encountered being \SI{57.6}{\ohm}
against $Z_{0}=\SI{50}{\ohm}$.

The minimax design is more exposed, as a design should be. Re-solving
\eqref{eq:obj} on the perturbed resonators moves the optimum over
$n^{2}\in[0.28,0.48]$ and $X_{s}\in[-90,\SI{-55}{\ohm}]$ and the achieved
worst-case sector reflection over $[0.248,0.330]$, that is $-12.1$ to
$\SI{-9.6}{\decibel}$. The predicted improvement on the as-built $-2.6$
\si{\decibel} is therefore \SI{8.5}{\decibel} nominally and at least
\SI{7.0}{\decibel} for any perturbation of magnitude $0.08$. The design point
itself is uncertain at the level quoted, and a converged unit cell should be
run before a geometry is committed; the qualitative conclusion, that a
probe alone is insufficient and that one series reactance suffices, is not.

\section{Consequences for the Radar Front End}
\label{sec:radar}

This section states what the active reflection of Section~\ref{sec:active}
and the unit-cell reflection of Section~\ref{sec:scan} mean at the module
that feeds the aperture. It is deliberately confined to two quantities the
antenna computation determines directly: the \emph{mismatch loss}, and the
\emph{power accepted by, and returned from, the radiating element}. Detection
range, range resolution, and any statement about a detection threshold depend
on the transmit power, the receiver noise figure, the losses between module
and element, the integration and the target model, none of which is analysed
here; converting the figures below into range would require that link budget
and this paper does not provide one. Nothing in this section is a new
electromagnetic computation.

\subsection{The element-level acceptance test is not a front-end test}
The quantity a module sees is the active reflection, not the isolated one.
With uniform amplitude the fraction of incident power accepted by port $n$ is
$1-|\Gamma_{n}|^{2}$, and the array average is $\etamm$ of \eqref{eq:etamm}.
Table~\ref{tab:frontend} collects the consequence. At broadside the aperture
accepts $0.780\pm0.087$ of the incident power and returns \SI{22.0}{\percent}
of it to the module, where the isolated interior element, at
$|S_{88}|=\SI{-16.94}{\decibel}$, returns \SI{2.0}{\percent}. The returned
power is larger by a factor of $10.9$, and the port-to-port spread of
$|\Gamma_{n}|$ from $0.350$ to $0.520$ puts the individual figures between
\SI{12.3}{\percent} and \SI{27.0}{\percent}.

Two things follow for the front end and neither is visible in an
element-level measurement. The mismatch loss is \SI{1.08}{\decibel} one way
and \SI{2.16}{\decibel} on a monostatic link where the same aperture
transmits and receives, against \SI{0.09}{\decibel} and
\SI{0.18}{\decibel} implied by the element figure. And the returned power
arrives at the circulator, the limiter and the output stage of the transmit
device rather than at the matched load an element test presents, which is a
rating and a thermal question rather than a loss. An acceptance criterion
written on the isolated $|S_{11}|$ certifies a module load that is wrong by
an order of magnitude in returned power.

\begin{table}[!t]
\caption{Power at the module. Accepted and returned fractions from the
computed reflection coefficients; mismatch loss only, no other link term}
\label{tab:frontend}
\centering
\footnotesize
\setlength{\tabcolsep}{3.5pt}
\begin{tabular}{lrrrr}
\toprule
& $|\Gamma|$ & accepted & returned & loss, 2-way \\
\midrule
\multicolumn{5}{l}{\emph{sixteen-element line, as built}}\\
isolated element & 0.142 & 0.980 & \SI{2.0}{\percent} & \SI{0.18}{\decibel} \\
active, broadside         & 0.469 & 0.780 & \SI{22.0}{\percent} & \SI{2.16}{\decibel} \\
active, \SI{30}{\degree}  & 0.275 & 0.925 & \SI{7.5}{\percent} & \SI{0.68}{\decibel} \\
active, \SI{60}{\degree}  & 0.315 & 0.901 & \SI{9.9}{\percent} & \SI{0.91}{\decibel} \\
worst port, broadside     & 0.520 & 0.730 & \SI{27.0}{\percent} & \SI{2.74}{\decibel} \\
\midrule
\multicolumn{5}{l}{\emph{$16\times24$ lattice, worst angle of the sector}}\\
as built                  & 0.741 & 0.452 & \SI{54.8}{\percent} & \SI{6.90}{\decibel} \\
minimax, predicted        & 0.279 & 0.922 & \SI{7.8}{\percent} & \SI{0.70}{\decibel} \\
\quad with the residual & 0.330 & 0.891 & \SI{10.9}{\percent} & \SI{1.00}{\decibel} \\
\bottomrule
\end{tabular}
\end{table}

\subsection{The module load moves with the beam, and is worst at broadside}
Because the active reflection is scan dependent, so is the load the module
drives. Across the sector examined the accepted fraction runs from $0.780$ at
broadside to $0.925$ at \SI{30}{\degree} and $0.901$ at \SI{60}{\degree}, a
mismatch-loss variation of \SI{0.74}{\decibel} one way. The extreme is at
broadside, which is the opposite of the usual expectation and follows from
the phase coincidence identified in Section~\ref{sec:active}: it is a
property of this lattice, substrate and feed, and Section~\ref{ssec:limits}
states when the condition fails.

For a design that sizes a circulator, a limiter or a load-pull contour to a
single operating point, the point to size to is therefore broadside, not the
sector edge. This is a statement about mismatch alone. It is not a statement
about sensitivity or about a detection threshold, both of which also involve
the embedded element gain of Section~\ref{sec:emb} and the rest of the link.

\subsection{What the matching designs would buy, if realised}
\emph{The designs of Section~\ref{sec:design} are circuit predictions on
full-wave extracted impedance. Neither has been re-solved in the full-wave
model, and the figures in this subsection inherit that status
(Section~\ref{ssec:tol}).} With that stated:

For the sixteen-element line, the re-matched probe is predicted to bring the
active reflection to $Z_{0}$ at \SI{9.35}{\giga\hertz} and to hold it below
\SI{-10}{\decibel} over \SI{459}{\mega\hertz}, that is an accepted fraction
above $0.99$ across that band, where the element as built does not reach
\SI{-10}{\decibel} active anywhere between 9.0 and \SI{9.7}{\giga\hertz}.
The \SI{152}{\mega\hertz} passive band of Section~\ref{sec:active} is a
manufacturing acceptance figure and is not the band over which the module
sees a specified load.

For the $16\times24$ lattice, the sector minimax is predicted to raise the
accepted fraction at the worst angle of the sector from $0.452$ to $0.922$,
that is from \SI{54.8}{\percent} of the power returned to the module to
\SI{7.8}{\percent}. Propagating the unit-cell residual of
Section~\ref{ssec:ucres} through the design gives a worst case of $0.891$
accepted, so the predicted improvement in worst-case sector reflection is
\SI{8.5}{\decibel} nominally and at least \SI{7.0}{\decibel} over that
residual.

\section{Conclusion}
\label{sec:conc}

The impedance seen at the probe of a patch in a periodic environment
separates into a feed inductance, a transformer ratio set by the probe
position, and an array-loaded resonator. Recovering the resonator from a
single unit-cell solution at an arbitrary probe position characterises the
whole probe-position design space, within the decomposition and the
hypotheses H1--H6 of Section~\ref{ssec:feas}, and three closed-form results
then follow. The set of impedances reachable at a fixed frequency and scan
angle is a disk. An exact match by probe position and resonant length
therefore exists if and only if $R_{p}\geq Z_{0}+X_{p}^{2}/Z_{0}$, with
optimal detuning $X_{p}/Z_{0}$ independent of the resonator resistance, and
when that fails the best attainable reflection follows from the image of the
disk under the bilinear map; and when one series reactance is admitted,
minimising the worst reflection over a scan sector is a quasiconvex problem,
so bisection with a convex feasibility test returns the global optimum.

Applied to a finite sixteen-element X-band line, the criterion is satisfied,
and relocating the probe from $0.331L$ to $0.207L$ from the radiating edge is
predicted to open \SI{459}{\mega\hertz} of active bandwidth where none
existed. Applied to the $16\times24$ periodic lattice of the same element ---
a different structure, solved as a unit cell and never as a finite array ---
it is violated at every scan angle examined, and the sector minimax is
predicted to improve the worst-case reflection from $-2.6$ to
$\SI{-11.1}{\decibel}$, at least \SI{7.0}{\decibel} of that surviving the
unit-cell adaptation residual. Both designs are predictions of the extracted
circuit model; neither has been re-solved in the full-wave model, and that is
the first of the three limitations below. The array as built is well matched in
isolation and badly matched when driven, by \SI{10.4}{\decibel}, under the
assumptions of Section~\ref{sec:model}, because the self and coupled terms of
the scattering matrix are in phase to within \SI{5.3}{\degree}; for this
array and this sector, broadside is consequently the worst-matched angle and
the synthesised array gain rises with scan. The mechanism is stated as a
condition on the arguments of $\mathbf{S}$ rather than as a general property
of microstrip arrays, and whether it arises in another design is a question
about that design.

Two methodological results accompany the analysis. Five physical consistency
identities detect post-processing errors in embedded-pattern recombination
that no solver residual reports and that are invisible in the quantity being
inspected. And a discrepancy between two independent numerical methods should
be decomposed before it is attributed: of the \SI{6.08}{\decibel} divergence
observed here at \SI{60}{\degree} of scan, at most \SI{4.07}{\decibel} is
attributable to the array environment and \SI{2.08}{\decibel} arises from the
two isolated-element models differing.

At the module that feeds the aperture, the consequence is a power budget
rather than a pattern. The array accepts $0.780$ of the incident power at
broadside and returns \SI{22.0}{\percent} of it, against \SI{2.0}{\percent}
for the isolated element, so the returned power a transmit device drives is
larger by a factor of $10.9$ than an element-level acceptance test would
certify, and the mismatch loss is \SI{2.16}{\decibel} on a monostatic link
rather than \SI{0.18}{\decibel}. That load is scan dependent and is worst at
broadside, which is where a circulator, a limiter or a load-pull contour
should be sized. These are consequences of the computed reflection
coefficients alone; detection range, range resolution and any statement about
a detection threshold need the rest of the link budget and are not claimed
here.

Three limitations bound what has been shown. The matching designs are
circuit predictions on full-wave extracted impedance and have not been
re-solved in the full-wave model. The mesh convergence of the sixteen-element
solution rests on two levels at a linear refinement ratio of 1.22, which does
not support an error estimate. And no prototype was built: every quantity in
this paper is computed, so the propositions are verified against numerical
experiment and against a second numerical method, but not against
measurement. Of the three, the propositions themselves are the least exposed,
since they are statements about a circuit decomposition and are proved rather
than fitted; what an experiment would test is hypothesis~H1 and the accuracy
of the extraction, not the algebra. Full-wave verification of the two
designs, a three-level convergence study, and fabrication and measurement of
the re-matched element are the natural continuation, in that order.

\appendices

\section{Proofs of Theorem~\ref{thm:disk} and Proposition~\ref{prop:feas}}
\label{app:a}

\emph{Theorem~\ref{thm:disk}.} Three steps: the locus at fixed $n^{2}$, the
union over $n^{2}$, and uniqueness.

\emph{(i)} By H2, as $\delta$ ranges over $\mathbb{R}$ the quantity
$1/(1+j\delta)$ traces the set
$\{w:|w-\tfrac12|=\tfrac12\}\setminus\{0\}$. This is elementary: writing
$w=1/(1+j\delta)$ gives $1/w=1+j\delta$, so $\Rez(1/w)=1$, and
$\Rez(1/w)=\Rez\bar{w}/|w|^{2}=1$ is $|w|^{2}=\Rez w$, that is
$|w-\tfrac12|^{2}=\tfrac14$; the value $w=0$ is excluded because $1/w$ is
then undefined, and it is approached as $\delta\to\pm\infty$. Multiplying by
$n^{2}R_{p}>0$ and adding $jX_{p}$, which are a positive dilation and a
translation, gives for each fixed $n^{2}$
\begin{equation}
\left|\Zact-\left(\tfrac{n^{2}R_{p}}{2}+jX_{p}\right)\right|
=\tfrac{n^{2}R_{p}}{2},\qquad \Zact\neq jX_{p} .
\label{eq:circn}
\end{equation}
Every circle \eqref{eq:circn} has $jX_{p}$ as its excluded point, and all of
them are internally tangent there.

\emph{(ii)} Put $u=\Rez\Zact$ and $v=\Imz\Zact-X_{p}$, so \eqref{eq:circn}
reads $u^{2}+v^{2}=n^{2}R_{p}u$ with $(u,v)\neq(0,0)$. Necessity: since
$0<n^{2}\leq1$ and $u>0$ on \eqref{eq:circn} except at the excluded point, we
have $u^{2}+v^{2}\leq R_{p}u$, which is $|Z-(R_{p}/2+jX_{p})|\leq R_{p}/2$,
so $\mathcal{R}\subseteq\mathcal{D}\setminus\{jX_{p}\}$. Sufficiency: let
$Z\in\mathcal{D}$ with $Z\neq jX_{p}$, so $(u,v)\neq(0,0)$ and
$u^{2}+v^{2}\leq R_{p}u$, which forces $u>0$. Set
\begin{equation}
n^{2}=\frac{u^{2}+v^{2}}{R_{p}u},
\label{eq:n2inv}
\end{equation}
which lies in $(0,1]$ by that same inequality, and choose $\delta=-v/u$,
which is finite because $u>0$. Substituting into \eqref{eq:reim} returns $Z$.
Hence $\mathcal{R}=\mathcal{D}\setminus\{jX_{p}\}$. Equality in
\eqref{eq:n2inv} holds exactly on $\partial\mathcal{D}$, where $n^{2}=1$, and
$\mathcal{D}$ is the closure of $\mathcal{R}$ because $jX_{p}$ is a limit
point of \eqref{eq:circn}.

\emph{(iii)} The pair constructed in (ii) is the only one: $\delta$ is fixed
by $\delta=-v/u$ and then $n^{2}$ by $n^{2}R_{p}=u(1+\delta^{2})$. Distinct
$(n^{2},\delta)$ therefore give distinct $\Zact$, and the parametrisation of
$\mathcal{R}$ is injective. $\blacksquare$

\emph{Proposition~\ref{prop:feas}.} By Theorem~\ref{thm:disk}, an exact
match exists if and only if $Z_{0}\in\mathcal{D}$, that is
\begin{equation}
\left(Z_{0}-\tfrac{R_{p}}{2}\right)^{2}+X_{p}^{2}\leq\left(\tfrac{R_{p}}{2}\right)^{2}
\iff Z_{0}^{2}-R_{p}Z_{0}+X_{p}^{2}\leq0,
\end{equation}
and dividing by $Z_{0}>0$ gives \eqref{eq:feas}. For the solution, separate
\eqref{eq:decomp} under H2,
\begin{equation}
\Rez\Zact=\frac{n^{2}R_{p}}{1+\delta^{2}},\qquad
\Imz\Zact=X_{p}-\frac{n^{2}R_{p}\delta}{1+\delta^{2}} .
\label{eq:reim}
\end{equation}
Imposing $\Rez\Zact=Z_{0}$ gives $n^{2}R_{p}=Z_{0}(1+\delta^{2})$;
substituting that into $\Imz\Zact=0$ gives $X_{p}=Z_{0}\delta$, hence
$\delta^{\star}=X_{p}/Z_{0}$, and returning to the first relation gives
$n^{2\star}R_{p}=Z_{0}+X_{p}^{2}/Z_{0}$. The pair is the unique solution of
the two equations, and it is admissible exactly when
$n^{2\star}\leq1$, which is again \eqref{eq:feas}. $\blacksquare$

Two properties of $\delta^{\star}$ follow. It is positive, so under H4 the
resonance must lie \emph{below} the operating frequency, by
$f_{r}\delta^{\star}/2Q$ in the parametrisation of \eqref{eq:decomp}; the
sign is fixed by $X_{p}>0$ and is not a design choice. And it depends on
neither $R_{p}$ nor $Q$, so within H1--H6 the required offset of the
resonance is determined by the feed inductance and the reference impedance
alone. The independence is a property of the decomposition, not of the
physical element: a probe relocation that changed $L_{p}$ appreciably would
violate H4, and \eqref{eq:feassol} would then be a first-order statement
rather than an exact one.

\emph{Corollary~\ref{cor:best}.} A bilinear map carries circles to circles.
Writing $Z=Z_{0}(1+\Gamma)/(1-\Gamma)$ in $|Z-c|=\rho$ and collecting terms
in $|\Gamma|^{2}$, $\Rez\Gamma$ and the constant gives
$A|\Gamma|^{2}+2\Rez[\overline{B}\,\Gamma]+C=0$ with $A$, $B$, $C$ as
defined, whence the image circle has centre and radius \eqref{eq:image}. The
extreme values of $|\Gamma|$ over a disk are $|w_{0}|\pm r_{w}$, and the
minimum is clipped at zero when the origin lies inside, giving
\eqref{eq:gmin}. $\blacksquare$

\section{Proof of Proposition~\ref{prop:qc}}
\label{app:b}
Fix $\theta_{s}$ and write $Z=R+jX$ with
$R=n^{2}\Rez\Zres$ and $X=n^{2}\Imz\Zres+X_{p}+X_{s}$, both affine in
$(n^{2},X_{s})$. For $0<r<1$,
\begin{equation}
\left|\frac{Z-Z_{0}}{Z+Z_{0}}\right|\leq r
\iff |Z-Z_{0}|^{2}\leq r^{2}|Z+Z_{0}|^{2},
\end{equation}
which expands to \eqref{eq:disk}. Since $1-r^{2}>0$, the left-hand side of
\eqref{eq:disk} is a convex quadratic in $(R,X)$; dividing by $1-r^{2}$ puts
it in the form
\begin{equation}
\left(R-Z_{0}\frac{1+r^{2}}{1-r^{2}}\right)^{2}+X^{2}
\leq\left(Z_{0}\frac{2r}{1-r^{2}}\right)^{2},
\end{equation}
a closed disk in the impedance plane, which is convex. The preimage of a
convex set under an affine map is convex, so the sublevel set
$\{(n^{2},X_{s}):|\Gamma(\theta_{s})|\leq r\}$ is convex. The sublevel set of
$F$ at level $r$ is the intersection of these sets over $\theta_{s}\in\Theta$
and is therefore convex for every $r$, which is the definition of
quasiconvexity. A quasiconvex function has connected sublevel sets, hence no
strict local minimum that is not global, and its minimum is located by
bisection on $r$ with a convex feasibility test at each step \cite{boyd2004}.
$\blacksquare$

For $r\geq1$ the sublevel set is a half-plane or the whole plane and the
argument is unchanged. The bisection converges linearly in $r$; sixty
iterations reduce the bracket below \num{1e-18}, which is well inside the
uncertainty of $\Zres$ itself.

\section*{Acknowledgment}
The author thanks the colleagues who reviewed earlier drafts of this
manuscript; their comments led to the uncertainty budget of
Section~\ref{sec:unc} and to the decomposition of Table~\ref{tab:attr}.


\end{document}